\documentclass[%
 aip,
 jmp,%
 amsmath,amssymb,
 reprint,%
]{revtex4-1}

\usepackage{subfigure}% Include figure files
\usepackage{graphicx}% Include figure files
\graphicspath{{./}{./images/}{./images_all/}{./images_all/lowMs_dist/}}
\DeclareGraphicsExtensions{.eps}
\usepackage{dcolumn}% Align table columns on decimal point
\usepackage{bm}% bold math
\begin{document}

\preprint{AIP/123-QED}

\title{Energy dissipation and switching delay in spin-transfer torque switching of nanomagnets with low-saturation magnetization in the presence of thermal fluctuations }% Force line breaks with \\
%\thanks{Footnote to title of article.}

\author{Kuntal Roy}
\email{royk@vcu.edu.}
\author{Supriyo Bandyopadhyay}
\affiliation{Department of Electrical and Computer Engineering, Virginia Commonwealth University, Richmond, VA 23284, USA}
\author{Jayasimha Atulasimha}
\affiliation{Department of Mechanical and Nuclear Engineering, Virginia Commonwealth University, Richmond, VA 23284, USA}
\author{Kamaram Munira}
\author{Avik W. Ghosh}
\affiliation{Department of Electrical and Computer Engineering, University of Virginia, Charlottesville, VA 22904, USA}

\date{\today}% It is always \today, today,

\begin{abstract}
A common ploy to reduce the switching current and energy dissipation
in spin-transfer-torque driven magnetization switching of shape-anisotropic single-domain nanomagnets is to employ
magnets with low saturation magnetization $M_s$ and high shape-anisotropy. 
The high shape-anisotropy compensates for low $M_s$ to keep the static 
switching error rate constant.  However,
this ploy increases the switching delay, its  variance in the presence of thermal noise, and  the dynamic switching error rate. 
Using the stochastic Landau-Lifshitz-Gilbert equation with a random torque emulating thermal noise, we show that  
pumping some excess spin-polarized current into the nanomagnet during switching 
will keep the mean
switching delay and its variance constant as we reduce $M_s$, while still reducing the energy dissipation significantly. 
\end{abstract}

% Use showkeys class option if keyword

\maketitle

\section{\label{sec:introduction}Introduction}

\begin{figure}
\includegraphics[width=2.2in]{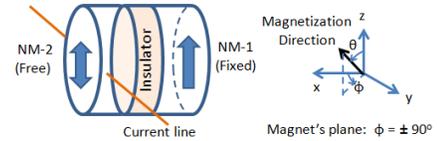}
\caption{\label{fig:MTJ} Simplified schematic diagram of an STT-RAM memory element. 
The nanomagnets (NMs) are on the $y$-$z$ plane and are shaped like 
elliptical cylinders. NM-1 is magnetically hardened along the $z$-axis so that
its magnetization direction is fixed. The magnetization direction of NM-2 can be rotated
with an in-plane spin polarized current that delivers a spin-transfer-torque. 
The magnetization orientation of the free layer NM-2 with respect to the z-axis 
($0^\circ$ and $180^\circ$)  encodes logic bits 0 and 1. }
\end{figure}

Spin-transfer-torque (STT) is an electric current-induced magnetization switching 
mechanism that can rotate the magnetization axis of a nanomagnet by exerting a torque on it due to the 
passage of a spin-polarized current~\cite{RefWorks:8,RefWorks:155,RefWorks:7,RefWorks:377}.  The STT-mechanism is routinely used to 
switch the magnetization of a shape-anisotropic nanomagnet from one stable orientation along the easy axis
to the other~\cite{RefWorks:300}, and has been demonstrated 
in numerous experiments involving both spin-valves~\cite{RefWorks:43} and  magnetic tunnel junctions 
(MTJs)~\cite{RefWorks:32}. MTJs, consisting of an insulating layer sandwiched between two ferromagnetic 
layers (one \emph{hard} and the other \emph{soft}), are becoming the staple of \emph{nonvolatile} magnetic random access 
memory (MRAM)~\cite{RefWorks:97,RefWorks:435} (see Fig.~\ref{fig:MTJ}). Switching the soft layer of an MTJ with the STT-mechanism
(STT-RAM) allows for high integration densities, but 
usually requires a high current density ($> 10^7$ A/cm$^2$) resulting in significant energy dissipation~\cite{roy11}.

 One way to decrease energy dissipation in STT-driven switching is to fashion nanomagnets out of
 materials with low saturation magnetization $M_s$ (e.g., dilute magnetic semiconductors). The spin-polarized switching current $I_s$, that delivers the spin-transfer-torque
and switches the magnetization, varies as $M_s^2$ (see Refs.~[\onlinecite{RefWorks:8,RefWorks:29,RefWorks:413,RefWorks:412}]), so that
the power dissipation $I_s^2R$ ($R$ = resistance of the nanomagnet) should vary as $M_s^4$ if $R$ does not change. However,
reducing $M_s$ decreases the in-plane shape anisotropy barrier $E_b$ that separates
 the two stable magnetization states along the easy axis.
This happens  because $E_b$ is proportional to the product of $M_s^2$ and the demagnetization
factor of the nanomagnet, which depends on the degree of shape anisotropy.
The decrease in $E_b$ increases the probability of random switching between 
the two stable states, which is $\sim$$\exp \left [-E_b/kT \right 
]$~\cite{RefWorks:348,RefWorks:148,RefWorks:436} at a temperature $T$. Therefore, if
one reduces $M_s$, then one must also increase the in-plane shape anisotropy (or aspect ratio of the
magnet) commensurately in order to keep the 
barrier $E_b$ and the static error probability  unchanged.
Increasing the shape anisotropy, or aspect ratio, has another beneficial effect; it  decreases the resistance $R$ in the path of the switching current $I_s$ if
the latter flows along the in-plane hard axis of the nanomagnet. This further 
reduces the power dissipation $I_s^2R$. 

It therefore appears that reducing $M_s$, while increasing shape anisotropy to
keep $E_b$ constant, is always beneficial.
  There is however one caveat. Reducing $M_s$ makes a nanomagnet more vulnerable to thermal 
  fluctuations~\cite{RefWorks:426} and can increase both the thermally averaged (mean) switching delay 
  and the standard deviation in the 
switching delay due to thermal fluctuations. This has a deleterious effect on
clock speed and clock synchronization in memory or logic technologies utilizing spin-transfer torque mechanism. Consequently, memory and logic devices utilizing materials with low saturation magnetization 
often work at low temperatures, even if the Curie temperature of the nanomagnet exceeds  
room temperature, just so that
 thermal agitations are suppressed~\cite{RefWorks:458,RefWorks:443}. 
In this paper, we show that a better approach to
contend with thermal fluctuations, when $M_s$ is reduced,
is to still work at room temperature, but use slightly more switching current 
than that required by the $I_s \propto M_s^2$ scaling law. This will decrease the mean switching delay and its variance, while still maintaining a significant energy saving due to the reduced $M_s$.

\section{Model}

We study the  magnetization dynamics of a nanomagnet
subjected to a spin-transfer-torque at room temperature by employing the stochastic
Landau-Lifshitz-Gilbert (LLG) equation~\cite{RefWorks:162,RefWorks:161}. It describes the time evolution
of the magnetization vector in the presence of spin-transfer-torque, the torque due to shape anisotropy, and 
an additional random torque due to thermal fluctuations. We choose the dimensions of the nanomagnet such that it has always a single ferromagnetic domain~\cite{RefWorks:133}. Thermal effects in
magnetization dynamics have been studied both theoretically~\cite{RefWorks:422,RefWorks:419,RefWorks:420,RefWorks:416,RefWorks:361,RefWorks:362} and experimentally~\cite{RefWorks:374,RefWorks:386,RefWorks:423}. 

We  consider a nanomagnet (see Fig.~\ref{fig:MTJ}) in the shape of an elliptical cylinder 
whose elliptical cross section lies in the $y$-$z$ plane with its major axis and minor axis aligned along the $z$-direction and the $y$-direction, 
respectively. The dimension of the major axis is $a$, that of the minor axis is $b$, and the thickness is $l$. The magnet's volume is $\Omega=(\pi/4)abl$. Let $\theta(t)$ be the angle subtended by the magnetization axis with the +$z$-axis at any instant of time $t$ and 
$\phi(t)$ be the angle between the +$x$-axis and the projection of the magnetization axis on the $x$-$y$ plane. Thus, $\theta(t)$ is the polar angle 
and $\phi(t)$ is the azimuthal angle.  Note that when 
$\phi$ = $\pm$ 90$^{\circ}$, the magnetization vector lies in the plane of the nanomagnet.

The potential energy of an isolated unperturbed shape-anisotropic single-domain nanomagnet is the uniaxial shape anisotropy energy given by 
\begin{equation}
E_{SHA}(t) = \frac{\mu_0}{2} M_s^2 \Omega N_d(t),
\label{eq:shape_anisotropy}
\end{equation}
\noindent
where $M_s$ is the saturation magnetization and $N_d$ is the demagnetization factor expressed as~\cite{RefWorks:157}
\begin{multline}
N_d(t) = N_{d-zz} cos^2\theta(t) + N_{d-yy} sin^2\theta(t) \, sin^2\phi(t) \\+ N_{d-xx} sin^2\theta(t) \, cos^2\phi(t)
\end{multline}
\noindent
with $N_{d-zz}$, $N_{d-yy}$, and $N_{d-xx}$ being the components of $N_d$ along the $z$-axis, $y$-axis, and $x$-axis,	respectively. The expressions for these quantities can be found in Ref.~[\onlinecite{RefWorks:402}] and they are constrained by the following relation:
\begin{equation}
N_{d-zz} + N_{d-yy} + N_{d-xx} = 1.
\end{equation}
We have assumed that the use of a properly balanced synthetic antiferromagnetic fixed layer can eliminate the net effect of dipole coupling on the free layer~\cite{RefWorks:302}.

At any instant of time, the total energy of the unperturbed isolated nanomagnet can be expressed as 
\begin{equation}
E(t) = E_{SHA}(t) = B(t) sin^2\theta(t) + C
\end{equation}
\noindent
where 
%\begin{subequations}
%\numparts
\begin{eqnarray}
B(t) &=& B(\phi(t)) = \frac{\mu_0}{2} \, M_s^2 \Omega \left\lbrack N_{d-xx} cos^2\phi(t) \right. \nonumber\\
&& \qquad\qquad \left. + N_{d-yy} sin^2\phi(t) - N_{d-zz}\right\rbrack, \label{eq:B}\\
C &=& \frac{\mu_0}{2} M_s^2 \Omega N_{d-zz}. \label{eq:C}
\end{eqnarray}
%\endnumparts
%\end{subequations}
\noindent
The in-plane shape anisotropy energy barrier height (using $\phi=\pm90^\circ$) can be expressed as
\begin{equation}
E_{SHA,in-plane} = \frac{\mu_0}{2} \, M_s^2 \Omega N_{d0}
\label{eq:N_d0}
\end{equation}
\noindent
where $N_{d0}=\left\lbrack N_{d-yy} - N_{d-zz}\right\rbrack$. Note that the in-plane shape anisotropy energy barrier height is independent of time $t$
even though $E_{SHA}(t)$ is not.

The magnetization \textbf{M}(t) of the single-domain nanomagnet has a constant magnitude at any given temperature but a variable direction, so that we can represent it by the vector of unit norm $\mathbf{n_m}(t) =\mathbf{M}(t)/|\mathbf{M}| = \mathbf{\hat{e}_r}$ where $\mathbf{\hat{e}_r}$ is the unit vector in the radial direction in spherical coordinate system represented by ($r$,$\theta$,$\phi$). The other two unit vectors in the spherical coordinate system are denoted by $\mathbf{\hat{e}_\theta}$ and $\mathbf{\hat{e}_\phi}$ for $\theta$ and $\phi$ rotations, respectively. The coordinates ($\theta$,$\phi$) completely describe the motion of \textbf{M}(t)~\cite{RefWorks:7}. 

The torque acting on the magnetization within unit volume due to shape anisotropy is 
\begin{eqnarray}
\mathbf{T_E} (t) &=& - \mathbf{n_m}(t) \times \nabla \mathbf{E}[\theta(t),\phi(t)] \nonumber\\
								 &=& - \{2 B(t) sin\theta(t) cos\theta(t)\} \mathbf{\hat{e}_\phi} - \{B_{0e}(t) \, sin\theta (t)\} \mathbf{\hat{e}_\theta}, \nonumber\\
\label{eq:torque_gradient}
\end{eqnarray}
\noindent
where
\begin{equation}
B_{0e}(t)=B_{0e}(\phi(t))=\frac{\mu_0}{2} \, M_s^2 \Omega (N_{d-xx}-N_{d-yy}) sin(2\phi(t)).
\label{eq:B_0e}
\end{equation}

Passage of a constant spin-polarized current $I_s$ through the nanomagnet generates a spin-transfer-torque that is given 
by~\cite{RefWorks:7}
\begin{equation}
\mathbf{T_{STT}}(t) = s  \, sin\theta(t) \, \mathbf{\hat{e}_\theta},
\label{STT-torque}
\end{equation}
\noindent
where $s = (\hbar/2e)\eta I_s$ is the spin angular momentum deposition per unit time and $\eta$ is the degree of spin-polarization in the current $I_s$. 

The effect of thermal fluctuations is to produce a \emph{random} magnetic field $\mathbf{h}(t)$  expressed as 
\begin{equation}
\mathbf{h}(t) = h_x(t)\,\mathbf{\hat{e}_x} + h_y(t)\,\mathbf{\hat{e}_y} + h_z(t)\,\mathbf{\hat{e}_z}
\end{equation}
\noindent
where $h_x(t)$, $h_y(t)$, and $h_z(t)$ are the three components in $x$-, $y$-, and $z$-direction, respectively. We will assume the same properties of the random field $\mathbf{h}(t)$ as described in the Ref.~[\onlinecite{RefWorks:186}]. Accordingly, the random thermal field can be expressed as~\cite{RefWorks:123}
\begin{equation}
h_i(t) = \sqrt{\cfrac{2 \alpha kT}{|\gamma| (1+\alpha^2) M_V \Delta t}} \; G_{(0,1)}(t) \qquad (i=x,y,z)
\label{eq:ht}
\end{equation}
\noindent
where $\alpha$ is the dimensionless phenomenological Gilbert damping constant, 
$\gamma = 2\mu_B \mu_0/\hbar$ is the gyromagnetic ratio for electrons  and is equal to
 $2.21\times 10^5$ (rad.m).(A.s)$^{-1}$, $\mu_B$ is the Bohr magneton, $M_V= \mu_0 M_s \Omega$, $1/\Delta t$ is the attempt frequency of the random thermal field affecting the magnetization dynamics; therefore $\Delta t$ should be chosen as the simulation time-step used to solve the coupled LLG equations numerically, and $G_{(0,1)}(t)$ is a Gaussian distribution 
with zero mean and unit standard deviation. Note that the variance in the random thermal fields is inversely proportional to the saturation magnetization $M_s$; therefore a lower saturation magnetization augments the detrimental effects of thermal fluctuations.

The thermal torque can be written as
\begin{equation}
\mathbf{T_{TH}}(t) = M_V \; \mathbf{n_m}(t) \times \mathbf{h}(t) = P_\theta(t)\,\mathbf{\hat{e}_\phi} - P_\phi(t)\,\mathbf{\hat{e}_\theta}
\end{equation}
\noindent
where
\begin{multline}
P_\theta(t) = M_V \lbrack h_x(t)\,cos\theta(t)\,cos\phi(t) + h_y(t)\,cos\theta(t)sin\phi(t) \\
 - h_z(t)\,sin\theta(t) \rbrack,
\label{eq:thermal_parts_theta}
\end{multline}
\begin{multline}
P_\phi(t) = M_V \lbrack h_y(t)\,cos\phi(t) -h_x(t)\,sin\phi(t) \rbrack.
\label{eq:thermal_parts_phi}
\end{multline}
\noindent

The magnetization dynamics of the single-domain nanomagnet under the action of various torques 
is described by the stochastic Landau-Lifshitz-Gilbert (LLG) equation as
\begin{multline}
\frac{d\mathbf{n_m}(t)}{dt} - \alpha \left(\mathbf{n_m}(t) \times \frac{d\mathbf{n_m}(t)}
{dt} \right)\\ = - \frac{|\gamma|}{M_V} \left\lbrack\mathbf{T_E}(t) + \mathbf{T_{STT}}(t) + \mathbf{T_{TH}}(t)\right\rbrack.
\label{LLG}
\end{multline}
\noindent

In spherical coordinate system, with constant magnitude of magnetization, we get the following coupled equations for $\theta$- and $\phi$-dynamics.
\begin{multline}
\left(1+\alpha^2 \right) \theta'(t) = \frac{|\gamma|}{M_V} \lbrack (B_{0e}(t) - s)
sin\theta(t) \\ - 2\alpha B(t) sin\theta (t)cos\theta (t) + \left(\alpha P_\theta(t) + P_\phi(t) \right) \rbrack\label{eq:theta_dynamics}
\end{multline}
\begin{multline}
\left(1+\alpha^2 \right) \phi'(t) = \frac{|\gamma|}{M_V} \lbrack \alpha (B_{0e}(t) - s) \\+ 2 B(t) cos\theta(t) - [sin\theta(t)]^{-1} \left(P_\theta(t) - \alpha P_\phi(t) \right)\rbrack.
\label{eq:phi_dynamics}
\end{multline}

The application of an in-plane spin-polarized current $I_s$ to produce spin-transfer torque results in an energy dissipation $I_s^2R\tau$, where 
$R$ is the in-plane resistance of the elliptical cylinder given by $R=\rho(2/\pi)\,(b/a)\,l$ with $\rho$ being is the resistivity of the material used and $\tau$ is the switching delay. 

Furthermore, because of Gilbert damping in the nanomagnet, an additional energy $E_d$ is dissipated when the 
magnetization axis in the nanomagnet switches from one orientation to the other along the easy axis. This energy is given by the expression $
E_d = \int_0^{\tau}P_d(t) dt$, where $\tau$ is the {\it switching delay} and $P_d(t)$ is the dissipated power given by 
\cite{RefWorks:319,RefWorks:124}
\begin{equation}
P_d(t) = \frac{\alpha \, |\gamma|}{(1+\alpha^2) M_V} \left| \mathbf{T_E}(t) + \mathbf{T_{STT}}(t)\right|^2.
\label{eq:Ed_dissipation}
\end{equation}
Thermal torque does not cause any net energy dissipation since mean of the thermal field is zero.

The thermal distributions of $\theta$ and $\phi$ in an unperturbed magnet are found 
by solving the Equations (\ref{eq:theta_dynamics}) and (\ref{eq:phi_dynamics}) while setting $I_s$ = 0.
This will yield the distribution of the magnetization vector's initial orientation ($\theta_{init}$, $\phi_{init}$) when stress is turned on. We consider magnetization intially situates at $\theta = 180^{\circ}$. The $\theta$-distribution is Boltzmann peaked at $\theta = 180^{\circ}$ with mean $\sim$175.5$^\circ$, while the $\phi$-distribution is Gaussian peaked at $\phi = \pm 90^{\circ}$~\cite{supplxd}.

The quantity $\tau$ for any switching trajectory is determined by solving the coupled equations (\ref{eq:theta_dynamics}) and (\ref{eq:phi_dynamics})
starting with an initial orientation ($\theta_{init}$, $\phi_{init}$) and terminating the trajectory when $\theta(t)$ reaches a pre-defined $\theta_{final}$,
regardless of what the corresponding $\phi_{final}$ is.  The time taken for a trajectory to complete 
(i.e., for $\theta(t)$ to reach $\theta_{final}$) is the value of $\tau$ for that trajectory. The average value $\langle \tau \rangle$ and the standard deviation $\langle \Delta \tau \rangle$
are found by simulating numerous (10,000) trajectories in the presence of the random thermal torque, and then extracting these quantities from the distribution.

The total energy $E_{total}$ dissipated in completing any trajectory is given by $E_{total} = E_d + I_s^2R \tau$. The average power dissipated 
in completing any trajectory is simply $E_{total}/\tau$. We can find the thermal average of $E_d$ and its variance by calculating $E_d$ for numerous trajectories and then computing these
quantities from the distribution.

\section{\label{sec:results}Simulation results}

\begin{figure}
\centering
\includegraphics[width=2.2in]{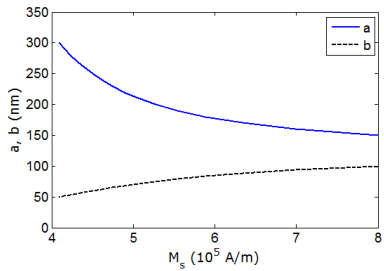}
\caption{\label{fig:low_Ms_a_b} Variation of the major axis ($a$) and the minor axis ($b$) with the saturation magnetization ($M_s$)  
needed to maintain a constant in-plane shape anisotropy barrier of 0.8 eV or $\sim$32 $kT$ at room temperature. The thickness of the nanomagnet is held constant at $l=2$ nm.}
\end{figure}
\begin{figure}
\centering
\includegraphics[width=2.2in]{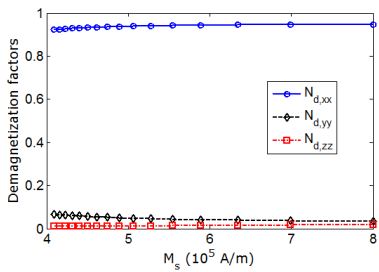}
\caption{\label{fig:demag_factors_Ms_cobalt} Demagnetization factors needed for different values of saturation magnetization $M_s$
in order to keep the in-plane shape anisotropy barrier constant at 0.8 eV or $\sim$32 $kT$ at room temperature. The demagnetization 
factors are computed from the major and minor axes values in Fig.~\ref{fig:low_Ms_a_b} and with constant thickness of 2 nm.} 
\end{figure}

We consider a nanomagnet made of CoFeB alloy which has low saturation magnetization~\cite{RefWorks:413} and a low Gilbert damping factor of 
$\alpha$ = 0.01. The saturation magnetization can be varied by varying the alloy composition~\cite{RefWorks:413}. The resistivity is assumed to be
the same as that of cobalt, i.e.
$\rho=5.81\times10^{-8}$ $\Omega$-m [Ref. \onlinecite{materials}]. We choose this material over dilute magnetic semiconductors which have
much smaller $M_s$ because the latter's $M_s$ is so small~\cite{RefWorks:458,RefWorks:443} that it will be impossible to make the in-plane shape anisotropy barrier $E_b$ (which is proportional to $M_s^2 \Omega N_{d0}$) large enough (32 $kT$ or 0.8 eV) without making the volume $\Omega$ of the nanomagnet very large.
With that large volume, the nanomagnet will no longer be single-domain. Choosing CoFeB 
 allows us to work at room temperature with a barrier of 0.8 eV or $\sim$32 kT, while
still ensuring single-domain behavior because the volume can be kept small.  

In order to maintain a constant value of $E_b$ = 0.8 eV
as we vary $M_s$, we increase the shape anisotropy of the
nanomagnet (or the aspect ratio $a/b$ of the ellipse) to increase $N_{d0}$ and 
compensate for any decrease in $M_s$. As we vary the aspect ratio $a/b$, 
we keep the cross-sectional area of the ellipse $[(\pi/4)\,ab]$ and the thickness $l$ constant, which keeps both the area and the volume of 
the nanomagnet $\Omega$ constant. The rationale behind keeping the cross-sectional area of the nanomagnet constant is to 
keep the density of devices per unit area on the chip constant. 

The in-plane shape anisotropy energy barrier depends on three quantities: $M_s$, $N_{d0}$ and $\Omega$ [see Equation~\eqref{eq:N_d0}]. 
Since we keep $\Omega$ constant, we compensate for any decrease in $M_s$ by commensurately increasing $N_{d0}$ alone. 
At all times, we ensure that the dimensions chosen ($a$, $b$, and $l$) guarantee that the nanomagnet remains in the 
single-domain limit~\cite{RefWorks:402,RefWorks:133}. The thickness $l$ is held constant at 2 nm.

Fig.~\ref{fig:low_Ms_a_b} shows how the major axis $a$ and the minor axis $b$ should vary with $M_s$
to keep the in-plane shape anisotropy energy barrier constant at 0.8 eV. This ensures that the \emph{static} error probability
associated with spontaneous switching between the two stable states along the easy axis remains constant as we 
vary $M_s$ and $N_{d0}$. In Fig.~\ref{fig:demag_factors_Ms_cobalt}, we plot the three components of  demagnetization factor
for different values of $M_s$ that will keep the in-plane shape anisotropy barrier constant at 0.8 eV. 
Obviously, as $M_s$ is decreased, we need to increase the value of $N_{d0}$, i.e. $\left(N_{d-yy}-N_{d-zz}\right)$, 
to keep the same in-plane shape anisotropy energy barrier height. With decreasing $M_s$, 
the quantity $N_{d-yy}$ increases significantly while $N_{d-zz}$ remains more or less constant. Since the three components of the demagnetization 
factor are constrained by the relation $N_{d-xx}+N_{d-yy}+N_{d-zz}=1$, the value of $N_{d-xx}$ must decrease proportionately, which
is seen in Fig.~\ref{fig:demag_factors_Ms_cobalt}. 

\subsection{Dependence of switching delay and energy dissipation on saturation magnetization}

We assume that when a spin-polarized current is applied to initiate switching, the magnetization vector
starts out from near the south pole ($\theta \simeq 180^\circ$) with a certain ($\theta_{init}$,$\phi_{init}$) picked from the initial angle distributions at 300 K~\cite{supplxd}. The magnetization dynamics ensures that $\theta$ continues to rotate towards $0^{\circ}$, while temporary backtracking of $\theta$ may occur due to random thermal kicks. Thermal fluctuations can introduce a spread in the time it takes to reach $\theta \simeq 0^\circ$ but cannot prevent magnetization to reach $\theta \simeq 0^\circ$. When $\theta$ becomes $ \leq 4.5^\circ$, switching is deemed to have completed. A moderately large number (10,000) of simulations, with their corresponding ($\theta_{init}$,$\phi_{init}$) picked from the initial angle distributions, are performed for each value of saturation magnetization to generate the simulation results in this subsection. The magnitude of the switching current $I_s$ is 2 mA at $M_s=8\times10^5$ A/m and it is reduced proportionately with the square of $M_s$ for other values of $M_s$. The spin polarization of the current is always 80\%. 

\begin{figure}
\centering
\includegraphics[width=2.2in]{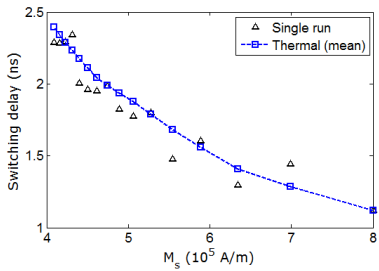}
\caption{\label{fig:thermal_delay_Ms_cobalt} Switching delay $\tau$ at room temperature (300 K) in a nanomagnet of fixed in-plane shape anisotropy energy barrier of 0.8 eV as a function of saturation magnetization $M_s$.  Switching current $I_s$ is 2 mA at $M_s=8\times10^5$ A/m and it is reduced proportionately with the square of $M_s$ for other values of $M_s$. The spin polarization of the current is always 80\%. }
\end{figure}

\begin{figure}
\centering
\includegraphics[width=2.2in]{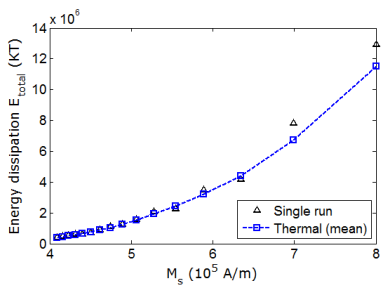}
\caption{\label{fig:thermal_energy_Ms_cobalt} Total energy dissipation in a nanomagnet of fixed in-plane shape anisotropy energy barrier 
of 0.8 eV as a function of saturation magnetization $M_s$ at room temperature (300 K). The total dissipation includes the dissipation in the 
switching circuit $I_s^2 R \tau$ and the internal energy dissipation $E_d$. The internal energy dissipation $E_d$ is only a very small fraction of the total energy dissipation $E_{total}$.}
\end{figure}

Figs.~\ref{fig:thermal_delay_Ms_cobalt} and~\ref{fig:thermal_energy_Ms_cobalt} show the mean switching delay and the mean
energy dissipation for different values of saturation magnetization $M_s$ when the in-plane shape anisotropy energy barrier is held constant at
0.8 eV by adjusting the nanomagnet's shape (and thus demagnetization factors) as we vary $M_s$. 
In generating these plots, the magnitude of the in-plane spin-polarized current is chosen as 2 mA when $M_s=8\times10^5$ A/m and the switching current $I_s$ was decreased in accordance with the $I_s \propto M_s^2$ scaling law.  Each scattered point (denoted `single run') in the 
figure is the 
switching delay for {\it one} representative switching trajectory at that value of $M_s$.
There is considerable scatter in the `single run' data which is significantly reduced by thermal
averaging (averaging over many trajectories).

The \emph{large} scatter in the `single run' data points is not caused by the random thermal torque since we get the similar trend without incorporating thermal fluctuations. The scatter is more prominent at smaller values of $M_s$, which corresponds to lower $I_s$ ($I_s \propto M_s^2$).
At lower $I_s$, the magnetization dynamics is more complex since there are more ripples (see Fig.~\ref{fig:dynamics_cobalt_Ms_4d09e5_Istt_523uA}
and Fig.~\ref{fig:dynamics_cobalt_Ms_4d09e5_Istt_1d05mA} later). As a result, there is more variability in the switching 
dynamics (and hence switching delay) with changing $I_s$ when the latter is small. This variability contributes to the scatter.

\begin{figure}
\centering
\includegraphics[width=2.2in]{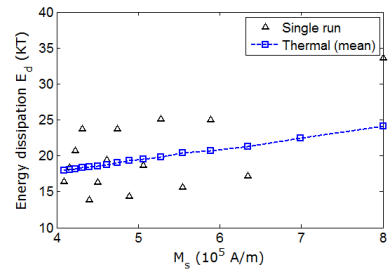}
\caption{\label{fig:thermal_energy_Ms_stt_and_shape_cobalt} Internal energy dissipation $E_d$ for different values of saturation magnetization at 300 K. This energy dissipation is several orders of magnitude smaller than the energy dissipation in the external circuitry thus it is only a very small fraction of the total energy dissipation $E_{total}$.}
\end{figure}

In Fig.~\ref{fig:thermal_delay_Ms_cobalt}, we find that the mean switching delay $\langle \tau \rangle$ decreases with  increasing saturation magnetization $M_s$.
This can be explained as follows. The spin-transfer torque is proportional to $I_s$/$M_s$. Since  $I_s \propto M_s^2$, the spin-transfer-torque becomes proportional to $M_s$. Therefore, reducing $M_s$ weakens the spin-transfer-torque. Since the in-plane shape anisotropy energy barrier is invariant, it takes longer for the weakened spin-transfer torque 
to overcome the in-plane shape anisotropy barrier and cause switching. This makes switching delay increase with decreasing $M_s$.

In Fig.~\ref{fig:thermal_energy_Ms_cobalt}, we plot the thermal means of the energy dissipation $E_{total}$ at 300 K as a function 
of the saturation magnetization $M_s$, while keeping the in-plane shape anisotropy barrier constant.
$E_{total}$ is overwhelmingly dominated by the component $I_s^2R \tau$, and the internal energy dissipation $E_d$ has a minor contribution [see Fig.~\ref{fig:thermal_energy_Ms_stt_and_shape_cobalt}]. 
The switching current $I_s$ varies as the square of $M_s$, 
so that $I_s^2$ varies as $M_s^4$. Furthermore, if we reduce $M_s$,
we have to increase the shape anisotropy (or the aspect ratio $a/b$) to keep the in-plane shape anisotropy energy barrier constant. If 
the switching current flows along the minor axis of the 
elliptical nanomagnet (always preferable since it results in minimum resistance in the path of the 
current), then increasing the ratio $a/b$ decreases the nanomagnet's 
electrical resistance proportionately. Thus, both $I_s$ and 
$R$ will decrease with decreasing $M_s$ (the latter because the in-plane shape anisotropy barrier is kept constant). Consequently,
 the {\it power} dissipation $I_s^2 R$ increases with $M_s$ 
 more rapidly than $M_s^4$. Unless the switching delay $\tau$ has a stronger dependence on $M_s$ than $\tau \propto M_s^{-4}$,
we will expect the energy dissipation to decrease with decreasing $M_s$ and that is precisely what we observe in 
Fig.~\ref{fig:thermal_energy_Ms_cobalt}.

The last two figures highlight two important facts: (1) the energy dissipated to switch can be reduced by decreasing $M_s$ while maintaining 
a fixed in-plane shape anisotropy energy barrier to keep the {\it static} error probability fixed, and (2) the switching delay increases 
if we reduce $M_s$ while keeping the in-plane shape anisotropy energy barrier fixed. Thus, there are {\it two} penalties involved with 
reducing energy dissipation by lowering $M_s$ and scaling $I_s$ quadratically with $M_s$: (i) slower switching, and (ii) higher {\it dynamic} error probability due to an increased variance in thermal field [see Equation (\ref{eq:ht})].

For the nanomagnet that we have considered (with the parameters described earlier), we find that lowering the saturation magnetization $M_s$ 
by a factor of $\sim$2 decreases the energy dissipation by $\sim$28 times while increasing the switching delay by approximately \emph{twice}. 
The factor of $\sim$2 decrease in $M_s$ causes a $\sim$16-fold decrease in $I_s^2$
(since $I_s \propto M_s^2$). Additionally, there is $\sim$4-fold decrease in the 
resistance of the nanomagnet owing to the fact that the shape anisotropy is increased to keep the in-plane shape anisotropy energy barrier constant. 
Thus, the $\sim$64-fold decrease in power dissipation and the 2-fold increase in switching delay together cause a net decrease 
of $\sim$28 times in the total energy dissipation. Therefore, if we decrease the
saturation magnetization by a factor of $\sim$2, then we will: (1) gain 28-fold in energy dissipation; (2) lose
2-fold in switching speed; and (3) lose somewhat in error rates due to thermal agitation since the variance in switching delay is increased.

\subsection{Constant switching delay scaling}

\begin{figure}
\centering
\subfigure[]{\label{fig:theta_dynamics_cobalt_Ms_8e5_Istt_2mA}\includegraphics[width=1.65in]
{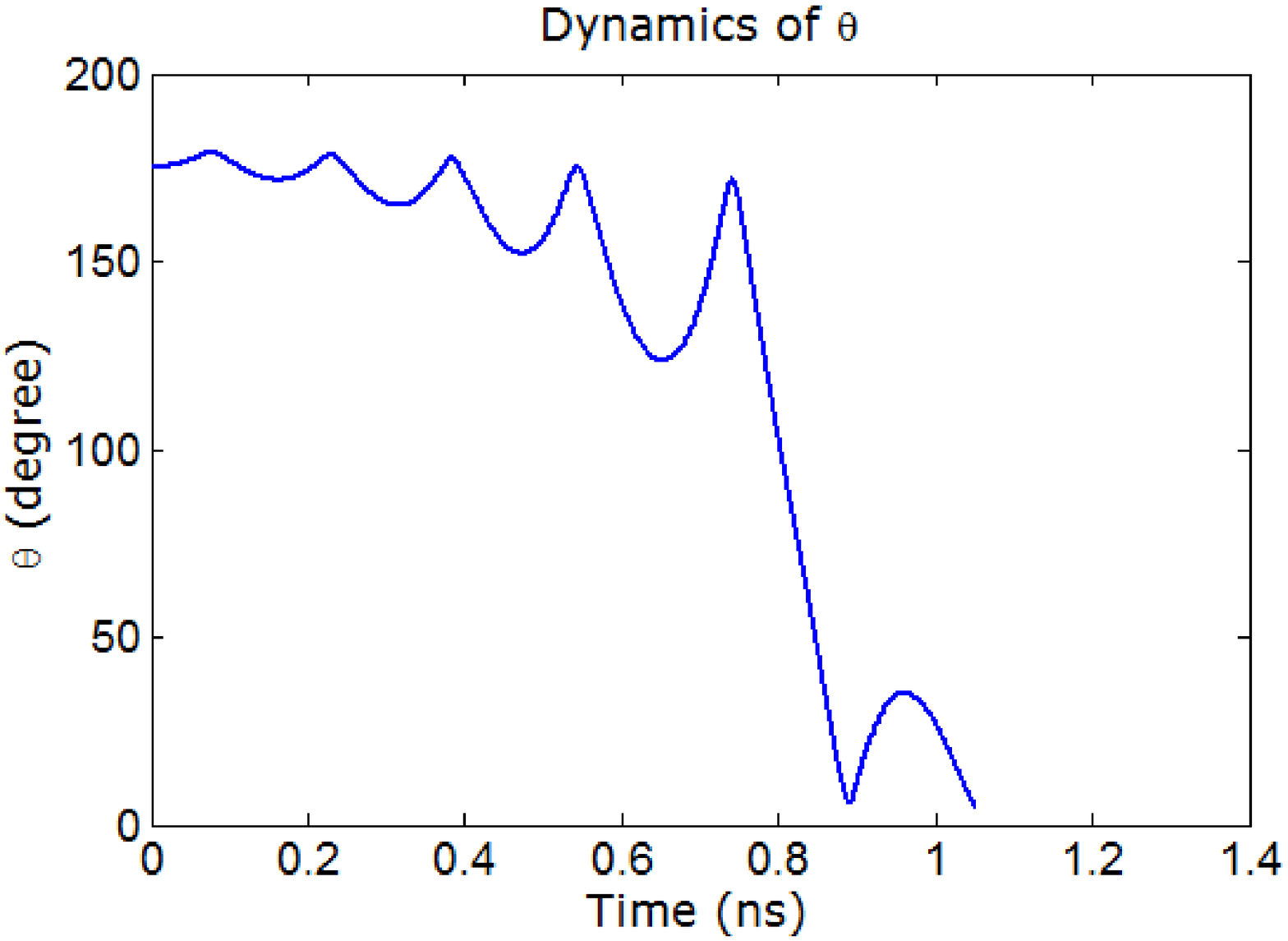}}
\subfigure[]{\label{fig:magnetization_dynamics_cobalt_Ms_8e5_Istt_2mA}\includegraphics[width=1.65in]
{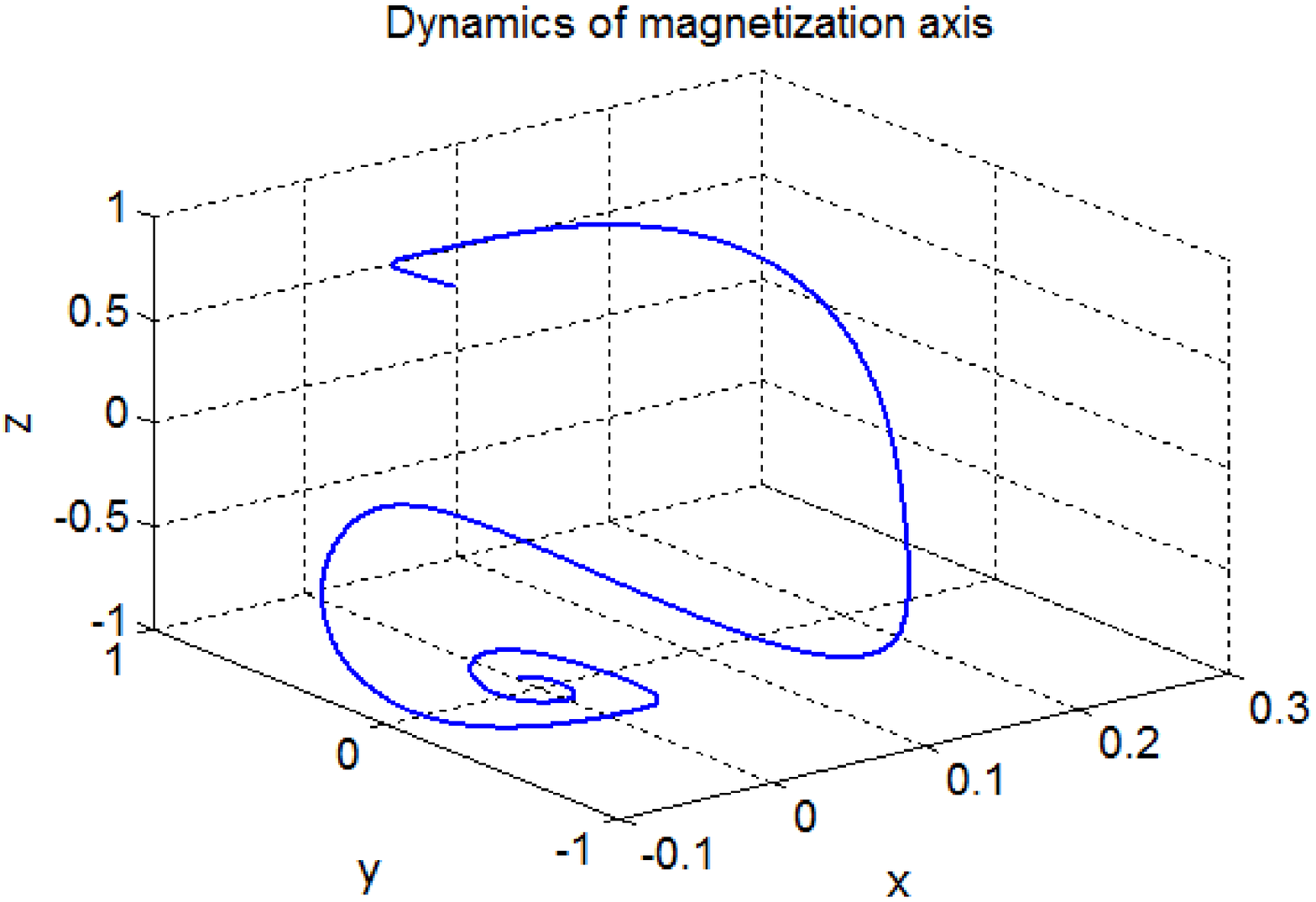}}
\caption{\label{fig:dynamics_cobalt_Ms_8e5_Istt_2mA} Switching dynamics of the magnetization vector in a nanomagnet of
major axis $a$ = 150 nm, minor axis $b$ = 100 nm, and $M_s = 8\times10^5$ A/m, and in-plane shape 
anisotropy energy barrier of 0.8 eV. This simulation does not consider any thermal fluctuations during the switching, however, the initial orientation of the magnetization is assumed to be $\theta_{init} = 175.5^{\circ}$ and $\phi_{init} = 90^{\circ}$ (thermally mean values for 300 K) to avoid the stagnation point exactly along the easy axis. Switching is caused by spin-transfer torque induced with
an in-plane current of 2 mA with 80\% spin polarization. (a) polar angle $\theta(t)$ versus time, 
and (b) the trajectory traced 
out by the tip of the magnetization vector during switching. Switching delay and energy dissipation are 
1.05 ns and $1.25\times10^7$ $kT$ [at room temperature], respectively.}
\end{figure}

\begin{figure}
\centering
\subfigure[]{\label{fig:theta_dynamics_cobalt_Ms_4d09e5_Istt_523uA}\includegraphics[width=1.65in]
{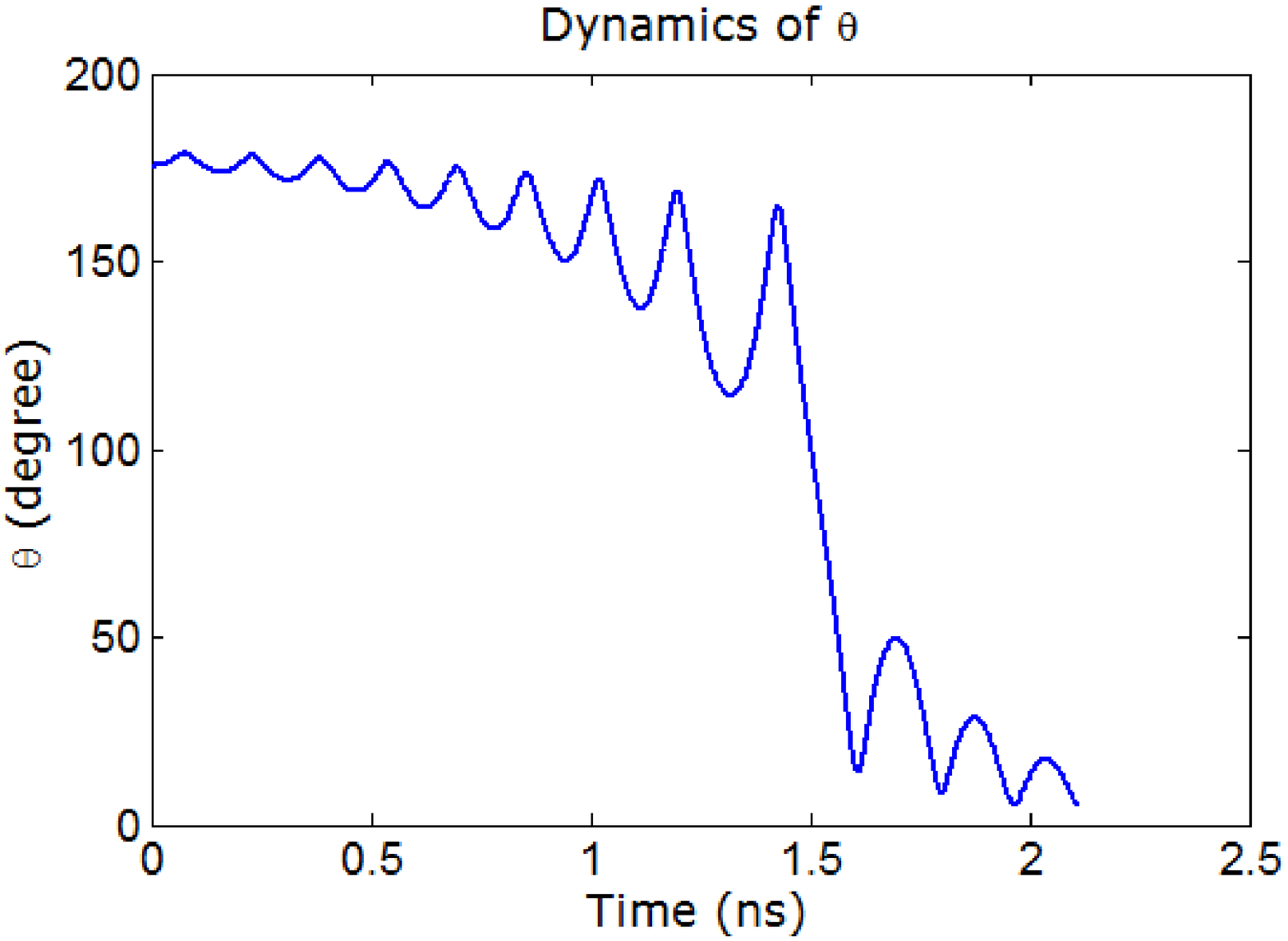}}
\subfigure[]{\label{fig:magnetization_dynamics_cobalt_Ms_4d09e5_Istt_523uA}\includegraphics[width=1.65in]
{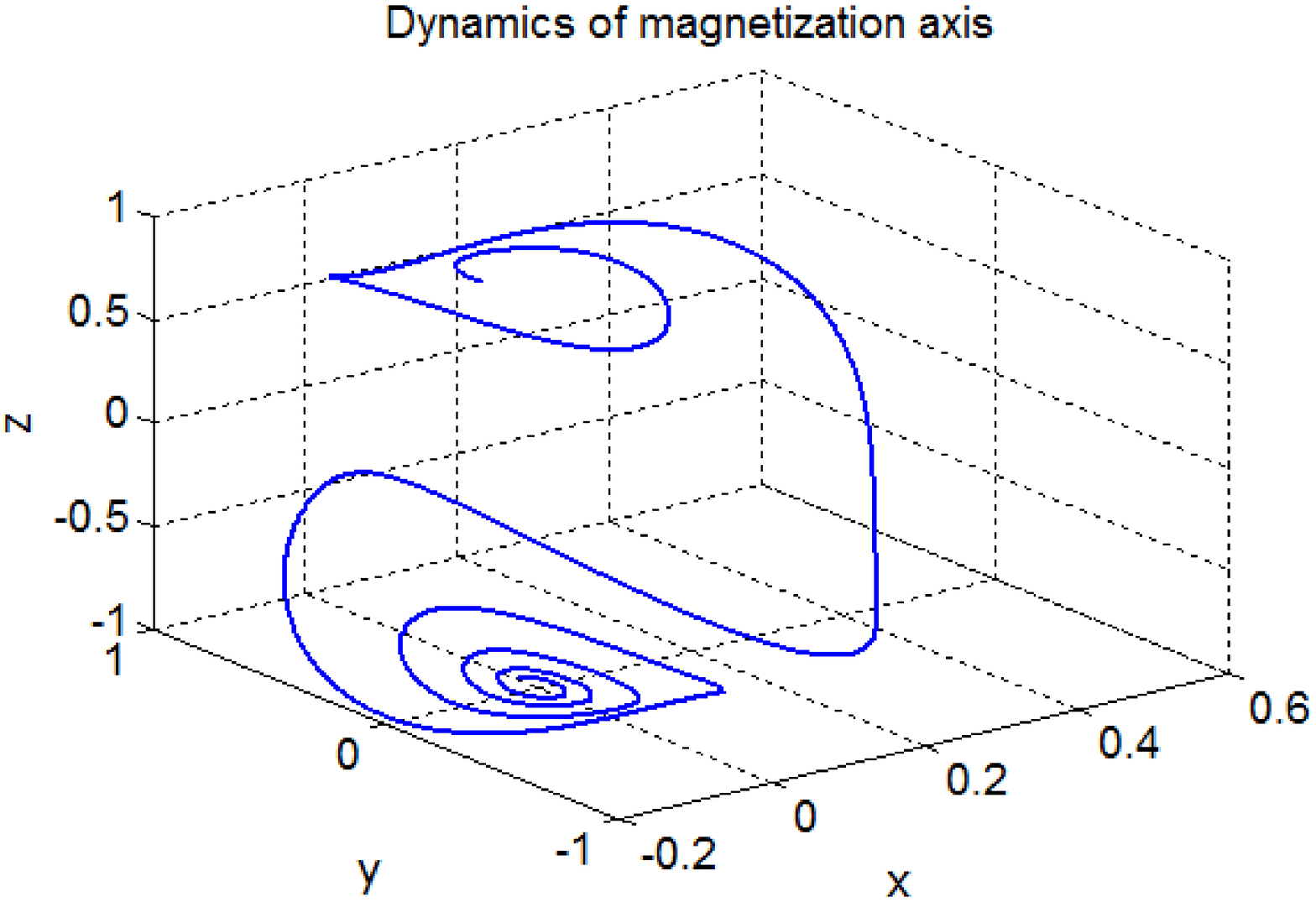}}
\caption{\label{fig:dynamics_cobalt_Ms_4d09e5_Istt_523uA} Switching dynamics for $M_s = 4.09\times10^5$ A/m and spin-transfer torque switching 
  current $I_s$ of 523 $\mu$A with 80\% spin polarization. 
  This simulation does not consider any thermal fluctuations during the switching, however, the initial orientation of the magnetization is assumed to be $\theta_{init} = 175.5^{\circ}$ and $\phi_{init} = 90^{\circ}$ (thermally mean values for 300 K) to avoid the stagnation point exactly along the easy axis. (a) Polar angle $\theta(t)$ versus time, and (b) 
  the trajectory traced 
out by the tip of the magnetization vector while switching occurs. 
The switching delay and the energy dissipation are 2.1 ns and $4.3\times10^5$ $kT$
[at room temperature], respectively.}
\end{figure}

\begin{figure}
\centering
\subfigure[]{\label{fig:theta_dynamics_cobalt_Ms_4d09e5_Istt_1d05mA}\includegraphics[width=1.65in]
{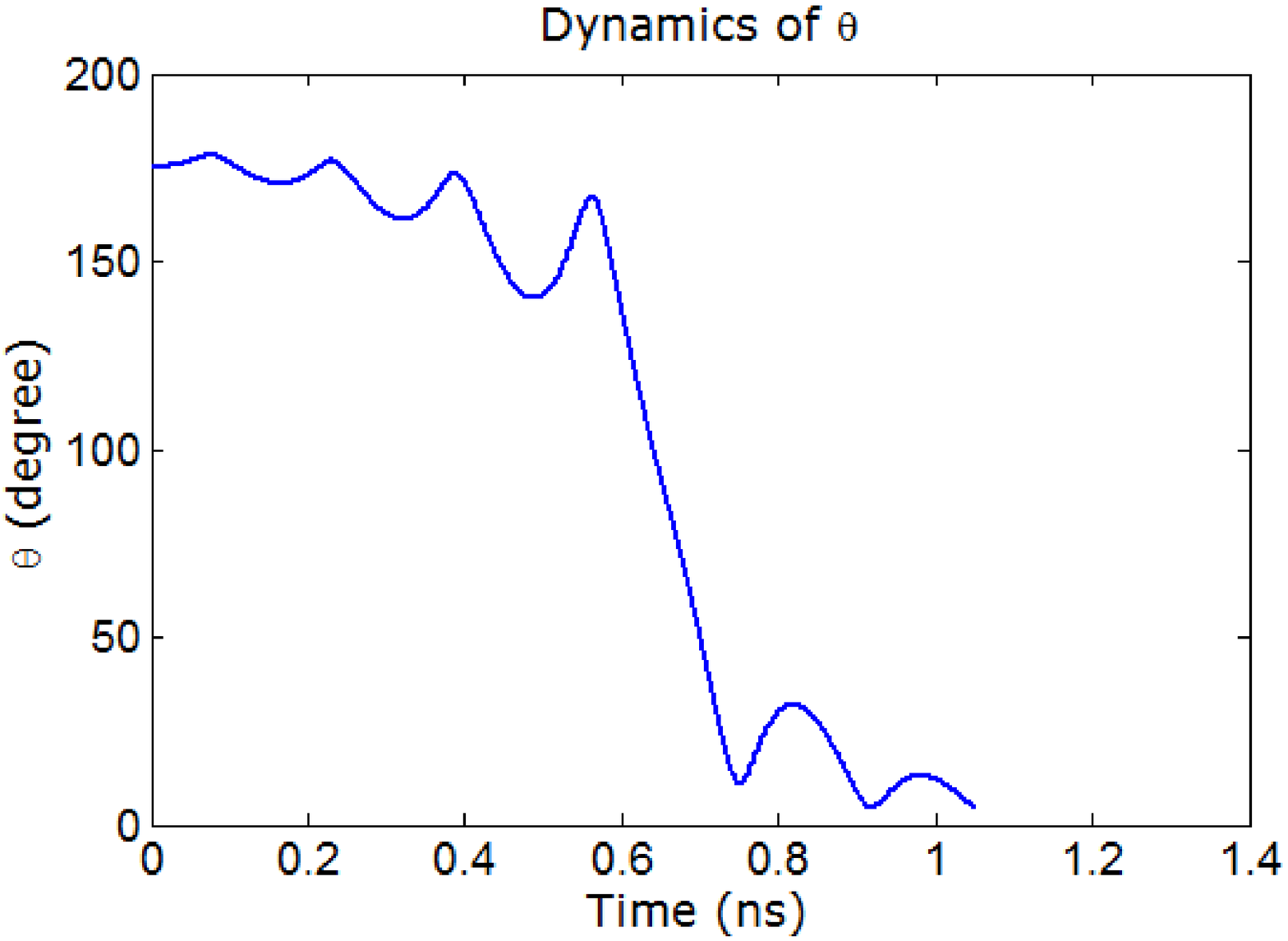}}
\subfigure[]{\label{fig:magnetization_dynamics_cobalt_Ms_4d09e5_Istt_1d05mA}\includegraphics[width=1.65in]
{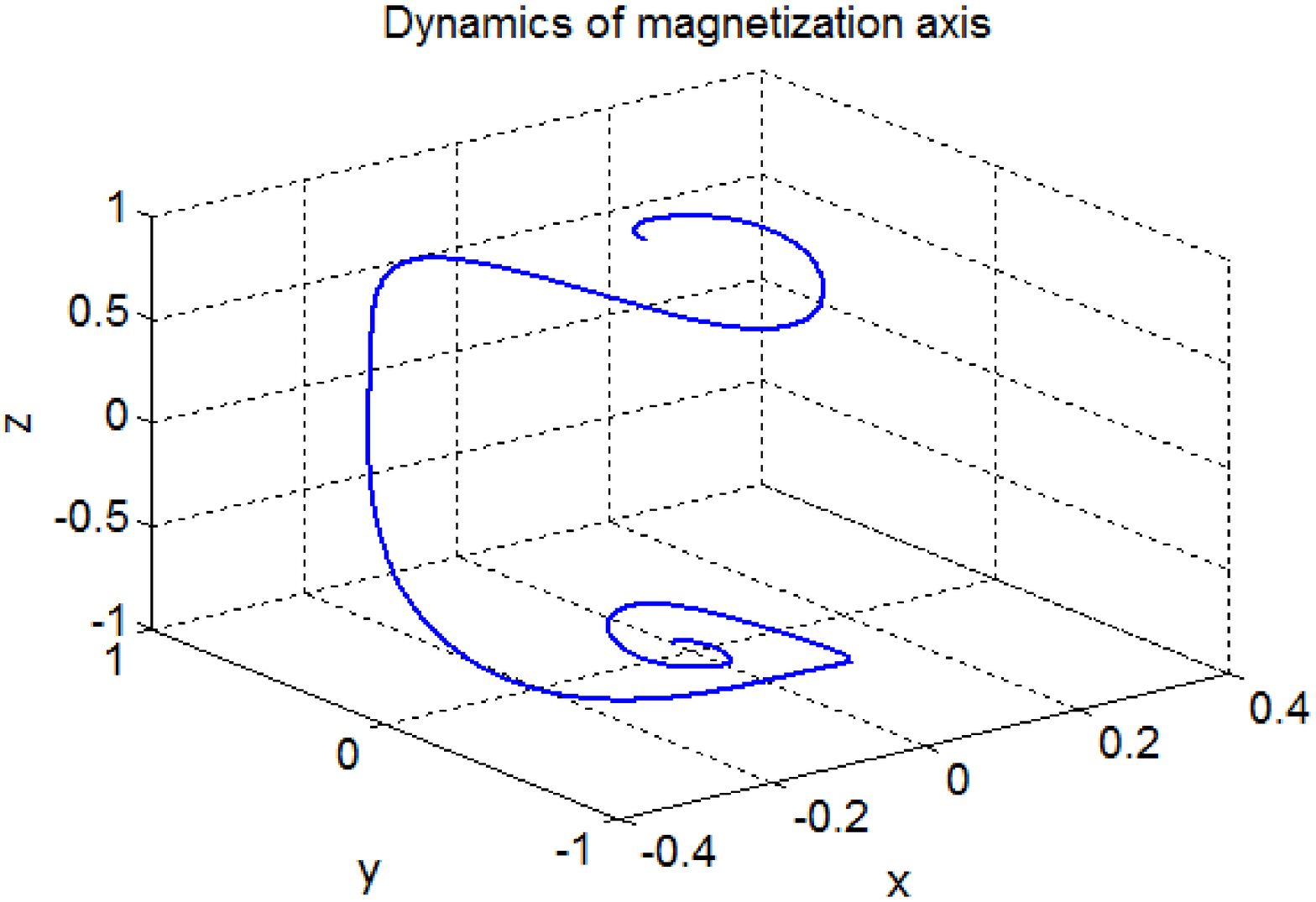}}
\caption{\label{fig:dynamics_cobalt_Ms_4d09e5_Istt_1d05mA} Switching dynamics for $M_s = 4.09\times10^5$ A/m and spin-transfer torque switching current $I_s$
of 1.05 mA with 80\% spin polarization. 
This simulation does not consider any thermal fluctuations during the switching, however, the initial orientation of the magnetization is assumed to be $\theta_{init} = 175.5^{\circ}$ and $\phi_{init} = 90^{\circ}$ (thermally mean values for 300 K) to avoid the stagnation point exactly along the easy axis. (a) Polar angle $\theta(t)$ versus time, and (b) the trajectory traced 
out by the tip of the magnetization vector while switching occurs. The switching delay and 
the energy dissipation are 1.05 ns and $8.6\times10^5$ $kT$ [at room temperature], respectively.}
\end{figure}

In order to understand how we can maintain a constant switching delay while scaling 
$M_s$, let us consider the relationship between switching current and switching delay.
In Figs.~\ref{fig:dynamics_cobalt_Ms_8e5_Istt_2mA} and~\ref{fig:dynamics_cobalt_Ms_4d09e5_Istt_523uA}, we plot the magnetization dynamics without considering any thermal fluctuations during the switching when $M_s = 8\times10^5$ A/m and $M_s = 4.09\times10^5$ A/m, respectively.
The switching current has been decreased from 2 mA for $M_s = 8\times10^5$ A/m to 523 $\mu$A for $M_s = 4.09\times10^5$ A/m, in accordance with 
the square-law scaling $I_s \propto M_s^2$.
In Figs.~\ref{fig:dynamics_cobalt_Ms_8e5_Istt_2mA} and~\ref{fig:dynamics_cobalt_Ms_4d09e5_Istt_523uA}, we have assumed the same 
initial orientation of magnetization $\theta_{init} = 175.5^{\circ}$ and $\phi_{init} = 90^{\circ}$, which are thermally mean values for 300 K to avoid the stagnation point exactly along the easy axis. The square-law scaling however results in an increased switching delay since the latter
 has obviously increased by a factor of 2 (from 1.05 ns to 2.1 ns). This has happened because of more ripples generating from more precessional motion of the magnetization vector seen in Fig.~\ref{fig:dynamics_cobalt_Ms_4d09e5_Istt_523uA}.
 In order to maintain the same switching delay
 of 1.05 ns as before, we will have to deviate from the square-law scaling 
and increase the switching current by nearly two times to 1.05 mA. 
Thus, we need to pump an excess current of 1.05 mA - 0.523 mA = 0.527 mA in order to maintain the same switching speed.
The corresponding magnetization dynamics without considering any thermal fluctuations during the switching is shown in  Fig.~\ref{fig:dynamics_cobalt_Ms_4d09e5_Istt_1d05mA}, where we 
have clearly recovered the 1.05 ns delay. 
The energy dissipation (dominated by $I_s^2 R \tau$) now goes up by a factor of two
[$I_s$ increases by a factor of two while  $\tau$ decreases by a factor of two]. 
Thus, we find that if we wish to maintain a {\it constant switching delay}, then 
we need to inject some excess current over that dictated by square-law scaling and therefore
suffer some excess energy dissipation. This excess energy dissipation is 
sufficiently small so that there is still considerable energy saving 
accruing from the reduction in $M_s$.
Reducing $M_s$ by a factor of $\sim$2 results
in a net energy saving of $\sim$14 times, instead of the $\sim$28 times estimated without imposing the requirement of constant switching
delay. The important point is that {\it we have extracted a very significant energy saving by reducing 
$M_s$ by a factor of 2, without sacrificing switching speed}.

\begin{figure}
\centering
\subfigure[]{\label{fig:theta_dynamics_cobalt_Ms_4d09e5_Istt_1d05mA_thermal}\includegraphics[width=1.65in]
{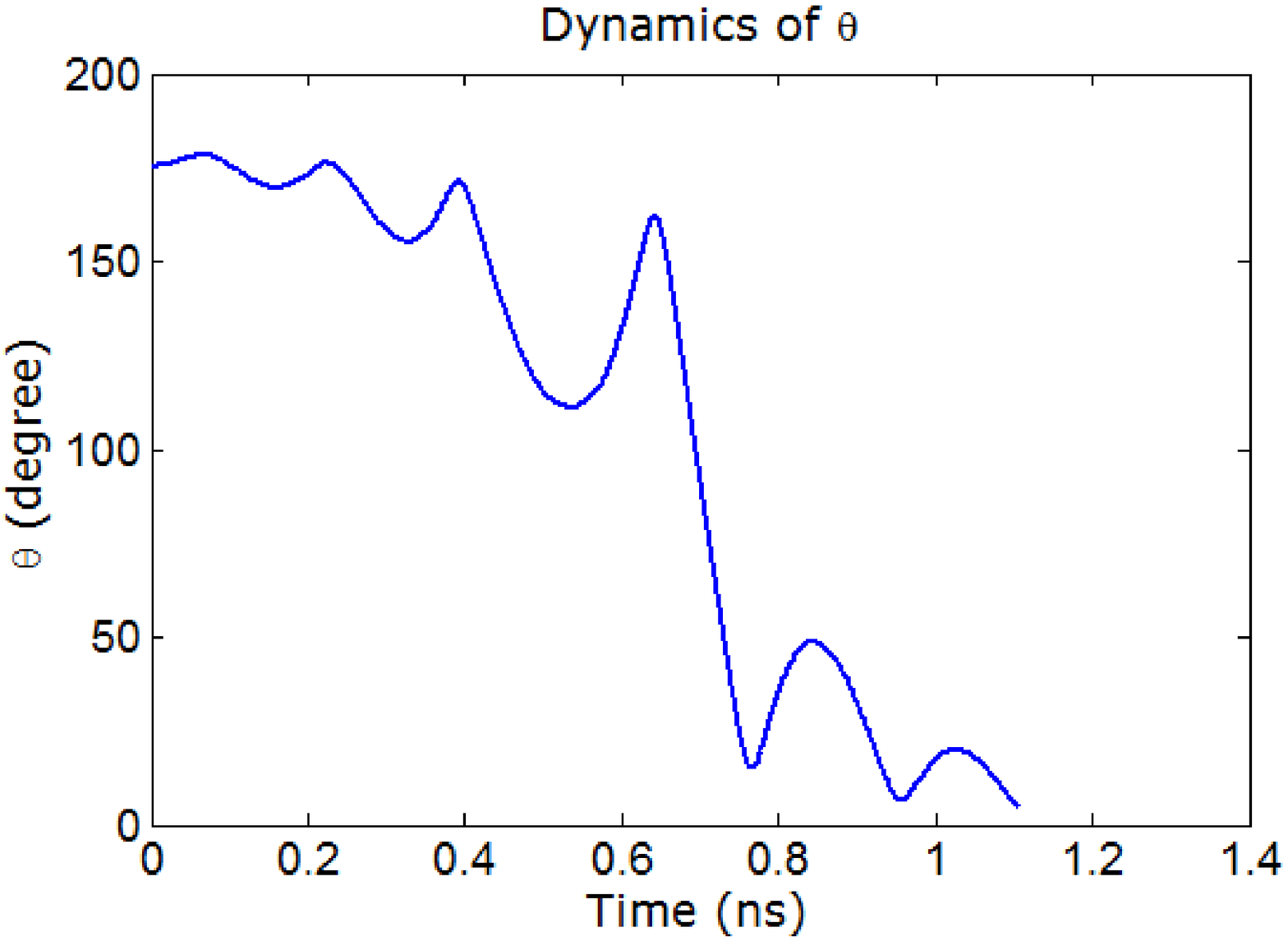}}
\subfigure[]{\label{fig:magnetization_dynamics_cobalt_Ms_4d09e5_Istt_1d05mA_thermal}\includegraphics[width=1.65in]
{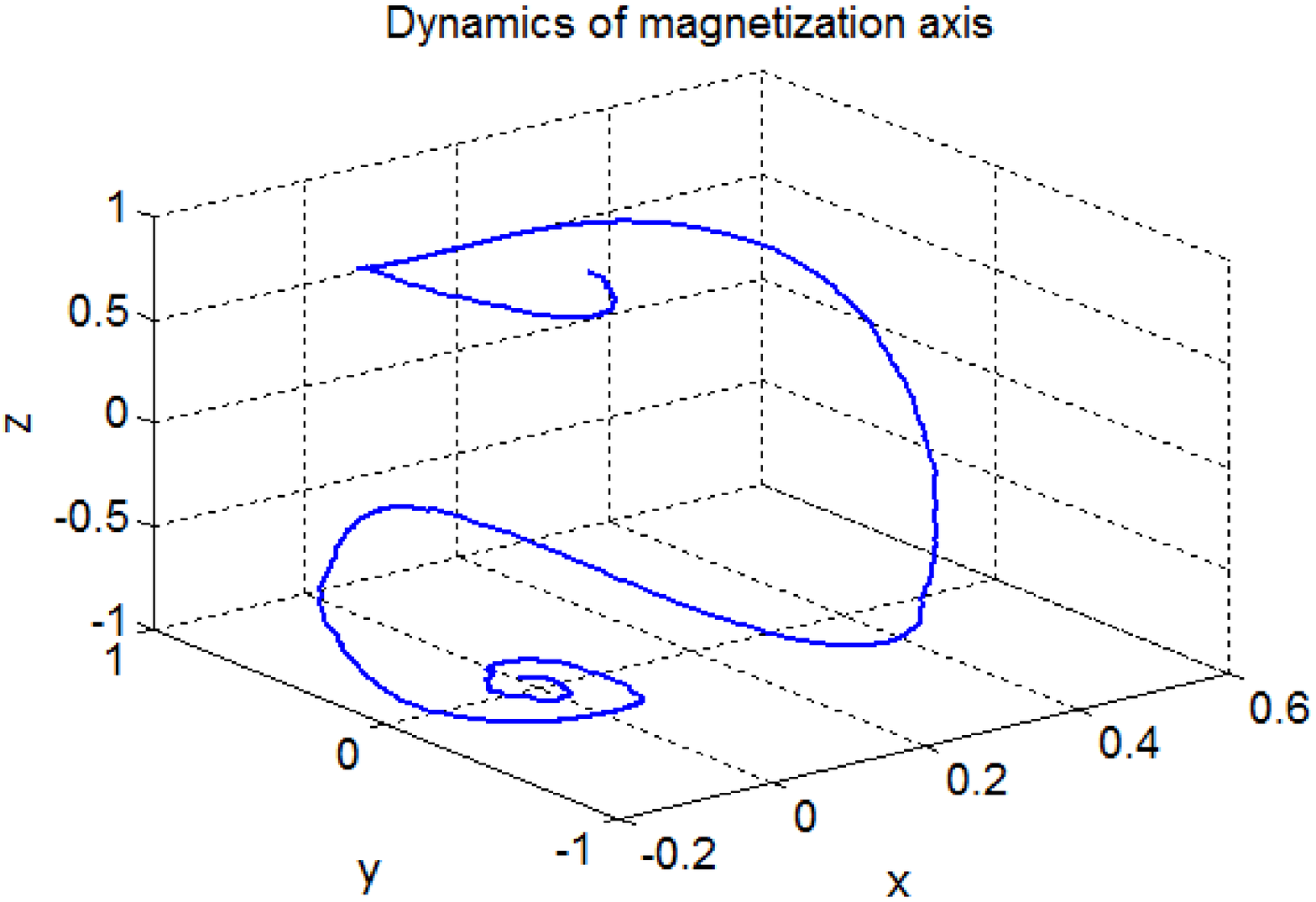}}
\caption{\label{fig:dynamics_cobalt_Ms_4d09e5_Istt_1d05mA_thermal} Switching dynamics at 300 K
for $M_s = 4.09\times10^5$ A/m and spin-transfer torque switching current $I_s$
of 1.05 mA with 80\% spin polarization. 
The initial 
orientation of the magnetization is  $\theta_{init} = 175.5^{\circ}$
and $\phi_{init} = 90^{\circ}$, which are the thermal mean values. This is one specific run from 10,000 MC simulations. (a) Polar angle $\theta(t)$ versus time, and (b) the trajectory traced out by the tip of the magnetization vector while switching occurs. The switching delay and 
the energy dissipation are 1.18 ns and $9.6\times10^5$ $kT$ [at room temperature], respectively.}
\end{figure}

\begin{figure}
\centering
\subfigure[]{\label{fig:thermal_delay_std_Ms_cobalt}\includegraphics[width=2.2in]
{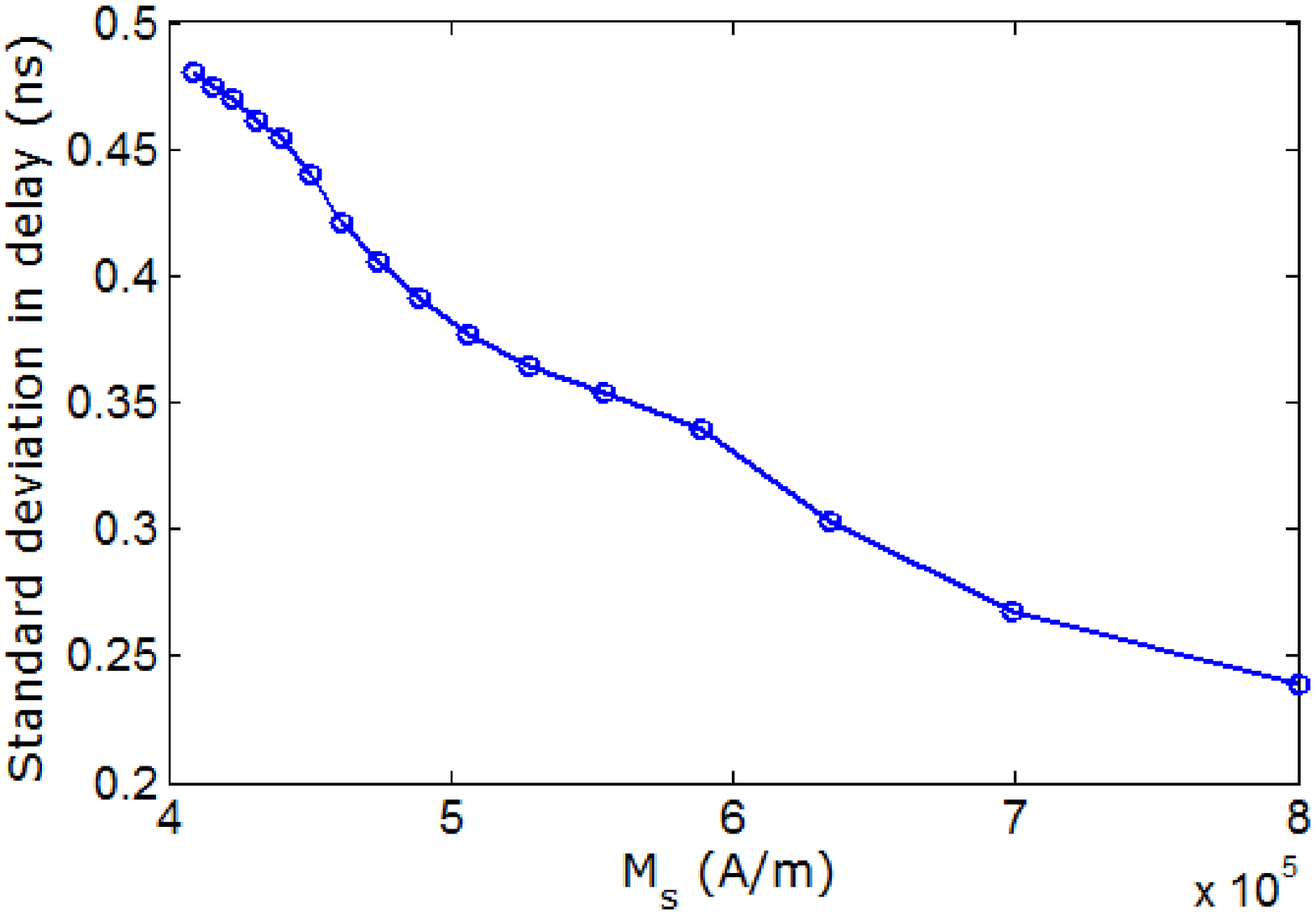}}
\subfigure[]{\label{fig:thermal_energy_std_Ms_cobalt}\includegraphics[width=2.2in]
{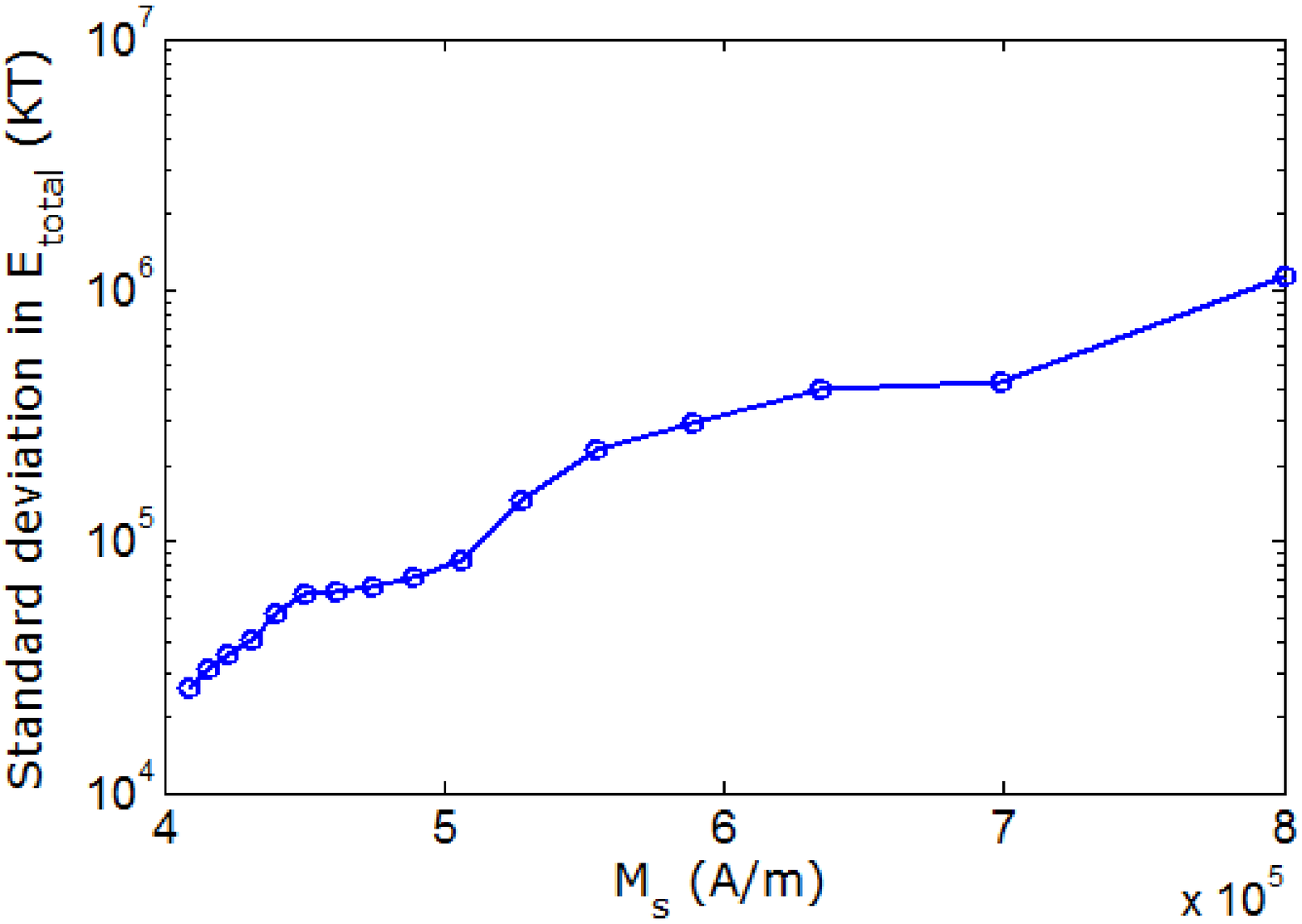}}
\caption{\label{fig:thermal_delay_energy_std_Ms_cobalt} Standard deviation in switching delay and energy dissipation due to thermal 
fluctuations at 300 K as a function of saturation magnetization $M_s$ in a nanomagnet with fixed in-plane shape anisotropy energy barrier of 
0.8 eV. (a) Standard deviation in switching delay. (b) Standard deviation in energy dissipation. }
\end{figure}

For illustrative purposes, we show in Fig.~\ref{fig:dynamics_cobalt_Ms_4d09e5_Istt_1d05mA_thermal} the
 magnetization dynamics in the presence of thermal fluctuations 
at 300 K for the same parameters as in Fig.~\ref{fig:dynamics_cobalt_Ms_4d09e5_Istt_1d05mA}. This is one representative run picked out from
10,000 simulations of the switching trajectory. Note that there is only some quantitative difference, but not much qualitative 
difference, between Figs.~\ref{fig:dynamics_cobalt_Ms_4d09e5_Istt_1d05mA} and~\ref{fig:dynamics_cobalt_Ms_4d09e5_Istt_1d05mA_thermal}.
The ripples are somewhat larger in amplitude and the precessional motion is slightly exacerbated. 
The switching delay has increased by $\sim$12\% in the presence of thermal agitations, however, it should be pointed out that the switching delay may decrease as well when the net effect of thermal agitations aids the magnetization rotation. 

Fig.~\ref{fig:thermal_delay_energy_std_Ms_cobalt} shows how the 
standard deviations in switching delay and energy dissipation due to thermal fluctuations depend on 
the saturation magnetization $M_s$. As expected, the standard deviation in switching delay 
increases with decreasing $M_s$, because the random thermal fields $h_i(t)$ (i=$x$,$y$,$z$), which are 
responsible for the standard deviation, has a $1/\sqrt{M_s}$ dependence [see Equation~\eqref{eq:ht}].
Furthermore, if we scale $I_s$ as $M_s^2$, then the spin-transfer torque also 
decreases as we reduce $M_s$ and that makes the increased thermal field even more 
effective in randomizing the switching delay.
For this reason, the error probability (or switching failure rate)
increases when $M_s$ decreases. This 
problem too can be overcome with some excess switching current. Our simulations 
have shown that if we increase the switching current from 523 $\mu$A to 1.05 mA, while 
holding $M_s$ constant at 4.09$\times$10$^5$ A/m, then the standard 
deviation in the switching delay goes down from 0.48 ns to 0.23 ns. Note that in this way we have recovered approximately the same standard deviation in switching delay as that for $M_s = 8\times$10$^5$ A/m.

 The standard deviation 
in energy dissipation however shows the opposite trend, i.e. it decreases with decreasing $M_s$. 
This happens because the
 energy dissipation $E_{total}$ is dominated 
by $I_s^2 R \tau$, which is proportional to $M_s^4 R \tau$. Lowering $M_s$ 
increases the standard deviation in $\tau$, but that increase is more than offset by the 
lower value of  $M_s$, so that the 
net standard deviation in $I_s^2 R \tau$ actually decreases with decreasing $M_s$. The excess
current that we pump now has a deleterious effect. If we increase the switching 
current from 523 $\mu$A to 1.05 mA while holding $M_s$ constant at 4.09$\times$10$^4$ 
A/m, then the standard deviation in energy dissipation goes up from 2.6$\times$10$^4$ kT
to 9.2$\times$10$^4$ kT.

The ratio of the standard deviation to the mean is relatively independent of $M_s$ for both the switching delay and the 
total energy dissipation at 300 K. This ratio does not vary by more than 5\% when $M_s$ is varied between 4.09$\times$10$^5$
and 8$\times$10$^5$ A/m.

\section{\label{sec:conclusions} Discussions and conclusions}

We have shown that one can significantly reduce energy dissipation in spin-transfer torque driven switching of shape-anisotropic 
nanomagnets by reducing the saturation magnetization of the magnet with appropriate 
material choice, while maintaining a constant in-plane shape anisotropy energy 
barrier (by increasing the magnet's aspect ratio) and a constant mean switching speed 
(by pumping some excess current) in the presence of thermal
fluctuations. Also, the increased variance in switching delay due to lower saturation magnetization can be mitigated by pumping some excess current. In the end, by employing these strategies, one can make the energy dissipation in 
spin-transfer torque driven switching of nanomagnets competitive with other technologies,
without sacrificing switching speed. This bodes well for 
applications of spin-transfer torque switched nanomagnets in non-volatile logic and memory.

%\bibliographystyle{aipnum4-1}
%%\nocite{*}
%\bibliography{royk,royk2}% Produces the bibliography via BibTeX.


\end{document}

\maketitle

In this supplementary section, we provide further details of the analysis and some additional results.

\tableofcontents

\section{Fluctuation about  the easy axis due to thermal torque}

We here explicitly show that thermal fluctuations by themselves can dislodge the magnetization from the easy axis. At T = 0 K, $\theta_{init} = 180^{\circ}$. In this case, spin-transfer-torque can never budge the magnetization vector
since it vanishes whenever $sin\,\theta = 0$. However, when T $>$ 0 K, thermal fluctuations can dislodge the magnetization from the easy axis.

In the spherical coordinate system, with constant magnitude of magnetization
\begin{equation}
\frac{d\mathbf{n_m}(t)}{dt} = \frac{d\theta(t)}{dt} \, \mathbf{\hat{e}_\theta} + sin \theta (t)\, \frac{d\phi(t)}{dt}
\,\mathbf{\hat{e}_\phi}.
\end{equation}
\noindent
Accordingly,
\begin{equation}
\alpha \left(\mathbf{n_m}(t) \times \frac{d\mathbf{n_m}(t)}{dt} \right) = - \alpha sin 
\theta(t) \, \phi'(t) \,\mathbf{\hat{e}_\theta} +  \alpha \theta '(t) \, \mathbf{\hat{e}_\phi}
\end{equation}
\noindent
and 
\begin{equation}
\frac{d\mathbf{n_m}(t)}{dt} - \alpha \left(\mathbf{n_m}(t) \times \frac{d\mathbf{n_m}(t)} {dt} \right) = \left\lbrack\theta '(t) + \alpha sin \theta (t)\, \phi'(t)\right\rbrack \, \mathbf{\hat{e}_\theta} + \left\lbrack sin \theta(t) \, \phi ' (t) - \alpha \theta '(t)\right\rbrack \,\mathbf{\hat{e}_\phi}
\end{equation}
\noindent
where $()'$ denotes $d()/dt$. Equating the $\hat{e}_\theta$ and $\hat{e}_\phi$ components of the LLG equation [Equation (16) in the main paper] and using Equations (8) and (10) in the main paper, we get
%\begin{subequations}
%\numparts
\begin{equation}
\theta ' (t) + \alpha sin \theta(t) \, \phi'(t)  =  \frac{|\gamma|}{M_V} \lbrack B_{0e}(t) \, sin\theta(t)  -s \, sin\theta(t)  + P_\phi(t)  \rbrack
\label{eq:LLG_separate_theta} 
\end{equation}
\begin{equation}
sin \theta (t) \, \phi '(t) - \alpha \theta '(t) = \frac{|\gamma|}{M_V} \lbrack 2 B(t) sin\theta(t) cos\theta(t) + s \, sin\theta(t) -  P_\theta(t)\rbrack.
\label{eq:LLG_separate_phi}
\end{equation}

When $sin\,\theta=0$, Equations~(\ref{eq:LLG_separate_theta}) and~(\ref{eq:LLG_separate_phi}) yield
\begin{eqnarray}
\theta ' (t) &=&  \cfrac{|\gamma|}{M_V}\, P_\phi(t) \label{eq:initial_p_phi}\\
\alpha \theta '(t) &=& \cfrac{|\gamma|}{M_V}\, P_\theta(t). \label{eq:initial_p_theta}
\end{eqnarray}
\noindent
Substituting for $P_\theta(t)$ and $P_\phi(t)$ from Equations (14) and (15) as in the main paper and using $\theta=180^\circ$, we get
\begin{equation}
\alpha h_x(t) sin\phi(t) - \alpha h_y(t) cos\phi(t) = h_x(t) cos\phi(t) + h_y(t) sin\phi(t)
\end{equation}
which yields
\begin{equation}
\phi(t) = tan^{-1} \left( \frac{\alpha h_y(t) + h_x(t)}{\alpha h_x(t)-h_y(t)} \right).
\label{eq:phi_t_thermal}
\end{equation}
Using this value of $\phi(t)$ in either Equation~(\ref{eq:initial_p_phi}) or Equation~(\ref{eq:initial_p_theta}), we get  
\begin{equation}
\theta'(t) = -|\gamma| \frac{h_x^2(t) + h_y^2(t)}{\sqrt{(\alpha h_y(t)+h_x(t))^2 + (\alpha h_x(t) - h_y(t))^2}}.
\end{equation}
\noindent
We can see  from the above equation that thermal torque can dislodge the magnetization vector from the easy axis 
because the time rate of change of $\theta(t)$, i.e. $\theta'(t)$,  is non-zero, even though $sin\,\theta = 0$. 
Note that the initial rate of deflection of
the magnetization vector from the 
easy axis does not depend 
on the component of the random magnetic field along the $z$-axis $\left [ h_z(t) \right ]$, 
which is a consequence of having the $z$-axis as the easy axis of the nanomagnet. 
However, as soon as the magnetization is deflected from the easy axis
($sin \theta \neq$ 0), all the three components of the random field would come into play. 

\section{Thermal distributions for $\theta_{init}$ and $\phi_{init}$}

\begin{figure}[htbp]
\centering
\includegraphics[width=3in]{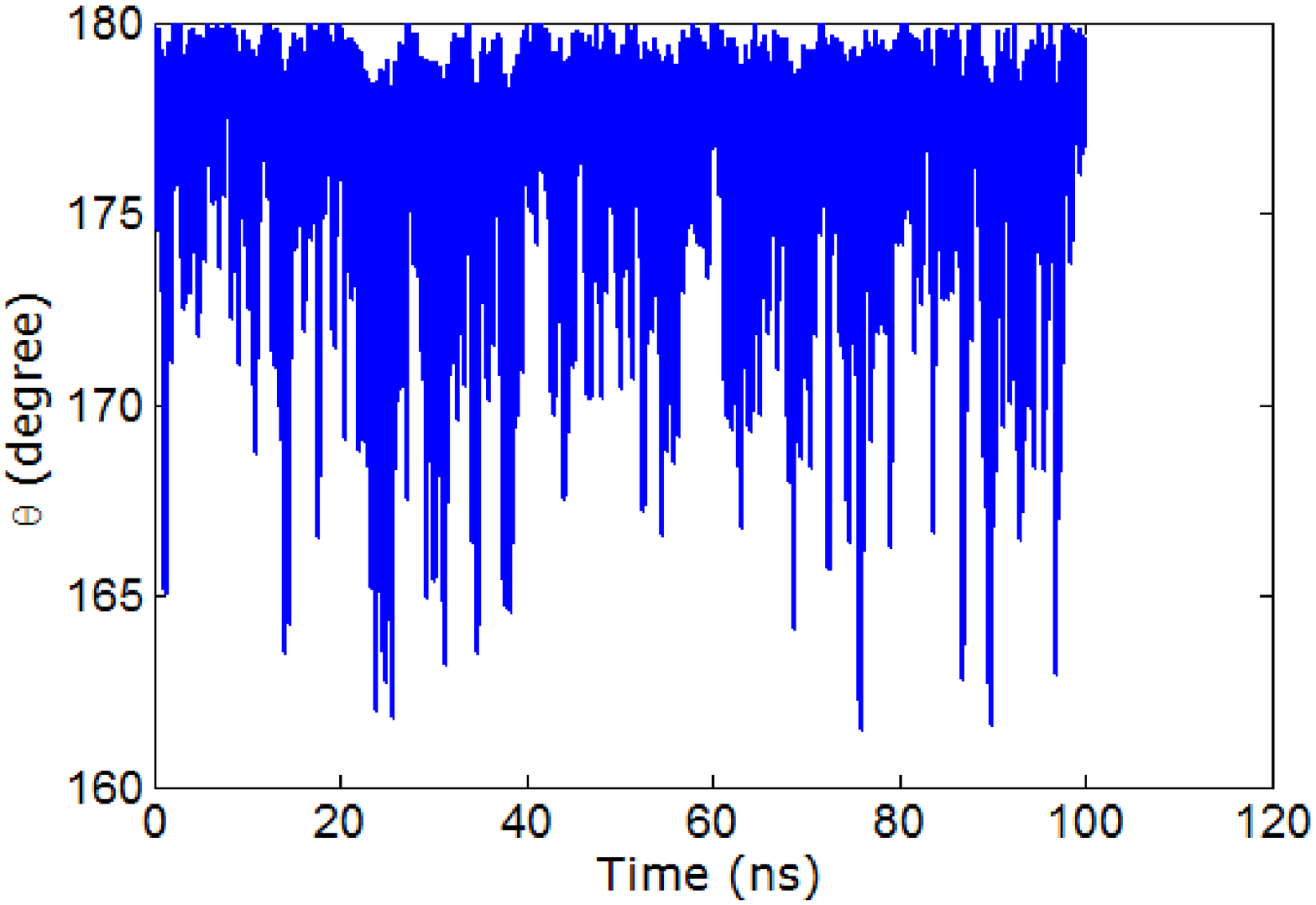}
\caption{\label{fig:theta_dynamics_thermal_cobalt_100ns} Temporal fluctuations in the polar angle $\theta(t)$  around one of the easy axes 
($\theta=180^\circ$) due to thermal torque at room temperature. The torque due to shape anisotropy is present since it is internal,
but there is no spin-transfer-torque which is always applied from an external source. Note that since $\theta$ is the polar 
angle, it is constrained to the interval [0$^{\circ}$, 180$^{\circ}$]. 
The nanomagnet has saturation magnetization $M_s=8\times10^5$ A/m, the major axis $a$ = 150 nm and the minor axis $b$ = 100 nm. 
The in-plane shape anisotropy energy barrier is 0.8 eV or $\sim$32 $kT$ at room temperature.}
\end{figure}

Fig.~\ref{fig:theta_dynamics_thermal_cobalt_100ns} shows the fluctuations in the polar angle
 $\theta(t)$ over time due to thermal agitations at 300 K, when the magnetization is nominally along 
 one of two stable orientations along the easy axis ($\theta$ = 180$^{\circ}$). There is no 
 external torque (e.g., spin-transfer torque) here, but there is a torque due to shape anisotropy 
 since it is a torque internal to the nanomagnet.

\begin{figure}[htbp]
\centering
\subfigure[]{\label{fig:thermal_distribution_cobalt_theta_100ns}\includegraphics[width=3in]
{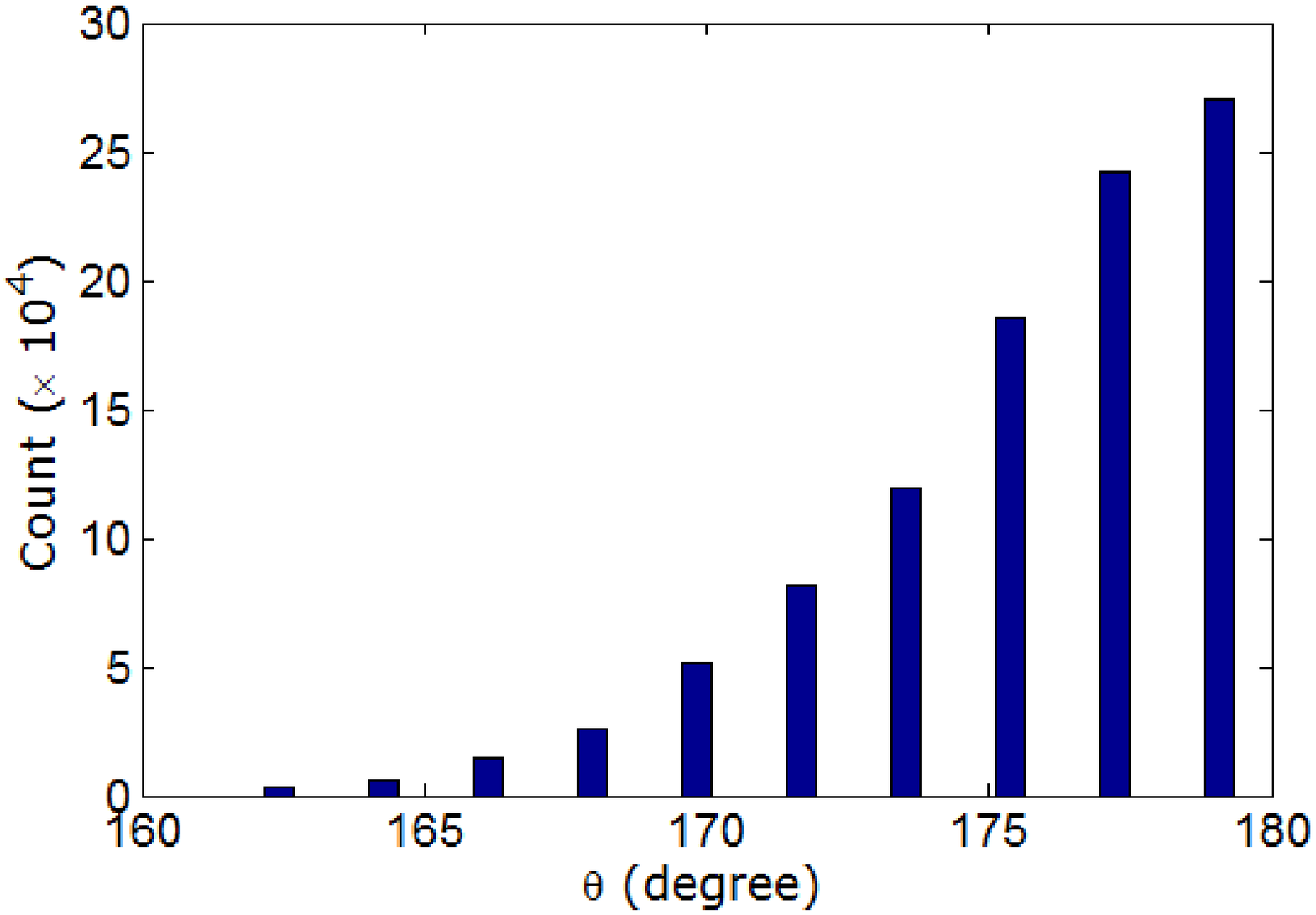}}
\subfigure[]{\label{fig:thermal_distribution_cobalt_phi_100ns}\includegraphics[width=3in]
{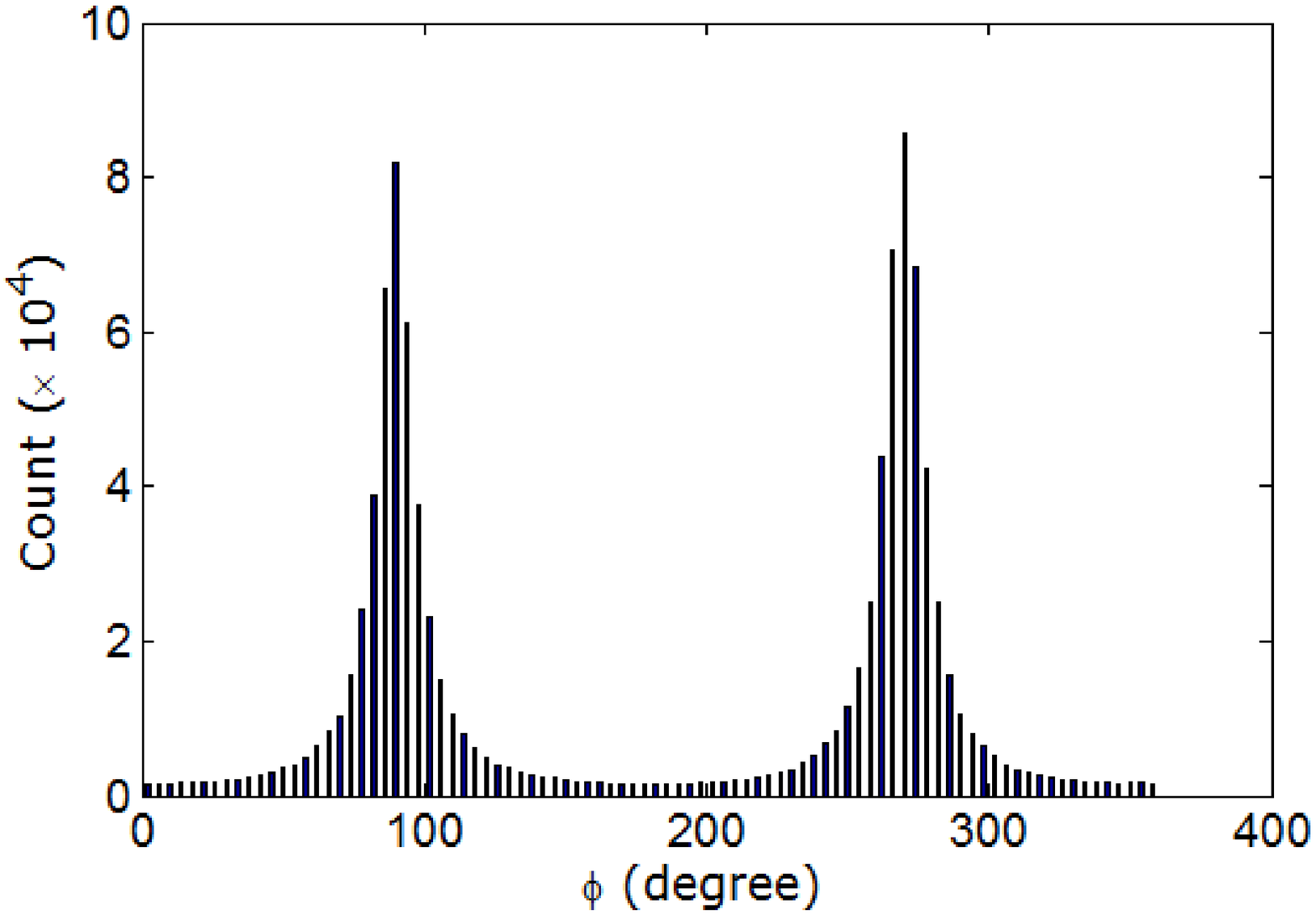}}
\caption{\label{fig:thermal_distribution_cobalt_100ns} Distributions of the polar angle $\theta(t)$ and azimuthal angle $\phi(t)$ due to thermal fluctuations at room temperature in a nanomagnet of elliptical cross-section. The major axis $a$ = 150 nm, the minor
axis $b$ = 100 nm and the saturation magnetization $M_s=8\times10^5$ A/m. The in-plane shape anisotropy energy barrier is 0.8 eV or $\sim$32 $kT$ at room temperature.
The magnetization vector is assumed to point  in one of two stable orientations along the easy axis ($\theta$ = 180$^{\circ}$)
but is perturbed by thermal fluctuations. (a) Distribution of polar angle $\theta(t)$ due to thermal fluctuations
at room temperature. The mean of the distribution is $\sim$$175.5^\circ$. (b) Distribution of azimuthal angle $\phi(t)$. 
There are two peaks in the distribution centered at $90^\circ$ and $270^\circ$ (or -90$^{\circ}$) that correspond to the plane of the nanomagnet.}
\end{figure}

In Fig.~\ref{fig:theta_dynamics_thermal_cobalt_100ns}, there are 1 million time steps in the 100 ns interval 
and this time interval is long enough to make the mean value of $\theta(t)$ independent of the interval 
duration. Fig.~\ref{fig:thermal_distribution_cobalt_100ns} shows the distributions of the polar 
angle $\theta(t)$ and the azimuthal angle $\phi(t)$ due to thermal fluctuations at 300 K within 
the same time interval of 100 ns used in Fig.~\ref{fig:theta_dynamics_thermal_cobalt_100ns}. 
The {\it most likely} value of $\theta$ is 180$^{\circ}$ which 
 is a stagnation point where spin-transfer-torque vanishes, but the mean value is not 180$^{\circ}$. The mean value is $\sim$$175.5^\circ$,
so that the 
time-averaged (mean)  deviation
 from the easy axis is $\sim$$4.5^\circ$. We noticed that halving the value of $M_s$ does not have any effect on the mean value of $\theta$, 
or the deviation from the easy axis, because the in-plane shape anisotropy energy barrier is kept constant  by adjusting the three components of 
demagnetization factor. Even though variance of the thermal field $h(t)$ depends on $M_s$, its mean value is zero, and hence
the mean value of the deviation from the easy axis turns out to be relatively independent of $M_s$.
The distribution of  $\theta(t)$ appears to be nearly \emph{exponential}, while the distribution of $\phi(t)$ contains 
two nearly \emph{Gaussian} distributions peaked at the two in-plane angles $\phi=\pm90^\circ$. 
Unlike the $\theta$-distribution, the $\phi$-distributions are {\it symmetric} about the mean values.

\section{Distributions of switching delay and energy dissipation in the presence of thermal fluctuations}

\begin{figure}[htbp]
\centering
\subfigure[]{\label{fig:thermal_distribution_delay_cobalt}\includegraphics[width=3in]
{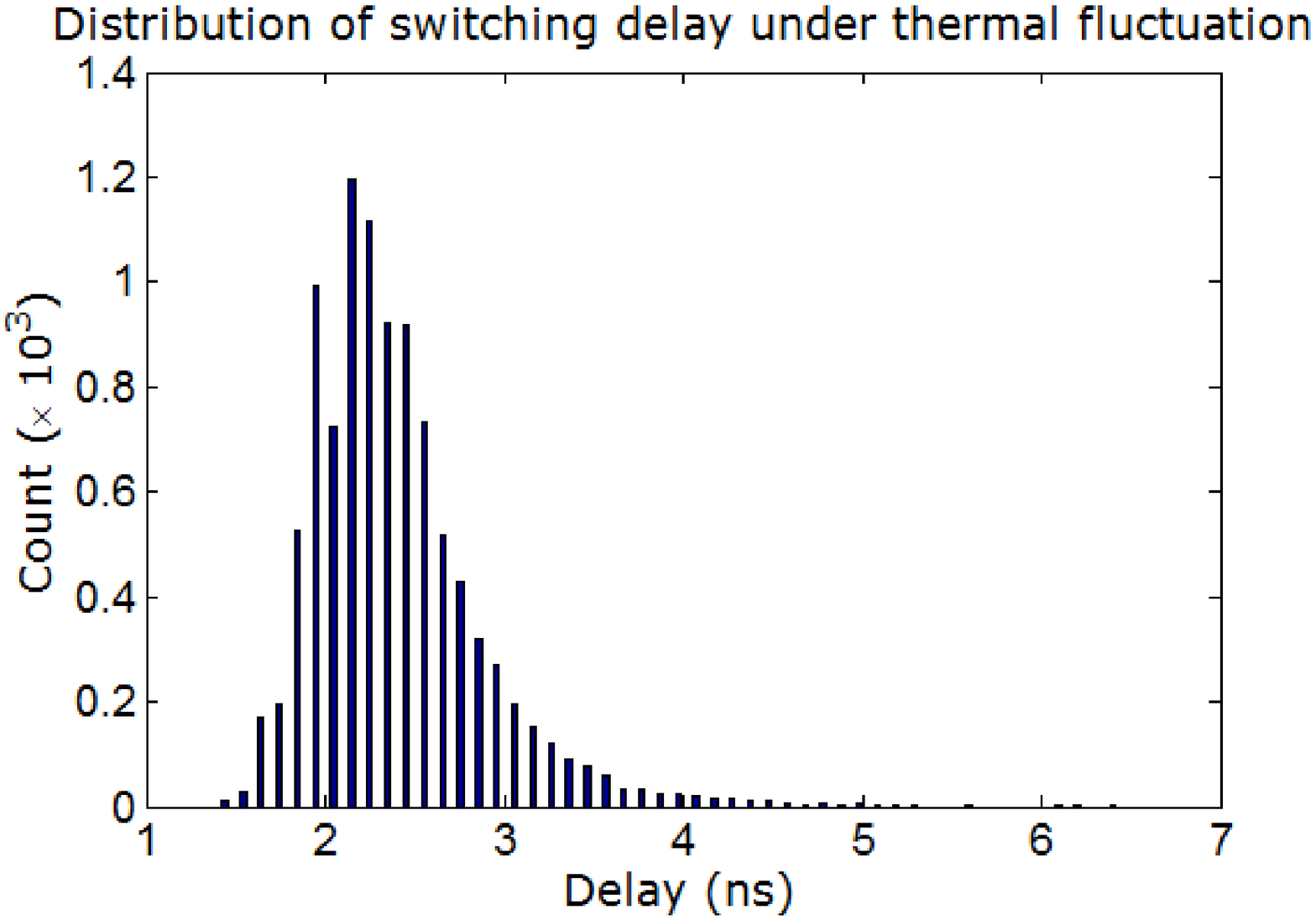}}
\subfigure[]{\label{fig:thermal_distribution_energy_cobalt}\includegraphics[width=3in]
{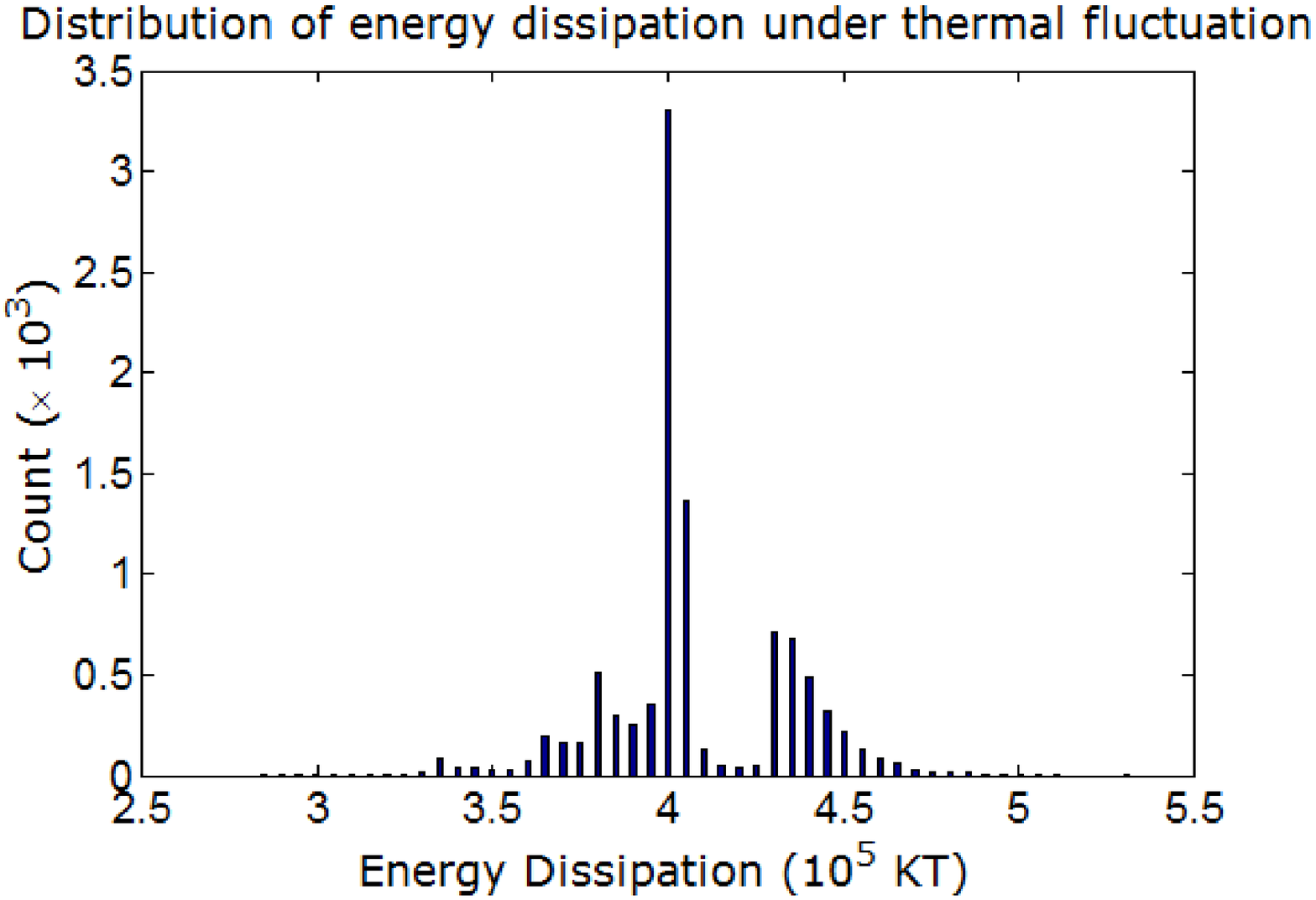}}
\caption{\label{fig:thermal_distribution_delay_energy_cobalt} Room temperature (300 K) 
distributions of switching delay and energy dissipation 
when a spin-transfer-torque 
is applied to switch a nanomagnet from one mean thermal deviation from one
orientation along the easy axis to one mean thermal deviation from the other diametrically opposite. Here,
the saturation magnetization $M_s=4.09\times10^5$ A/m and the in-plane shape anisotropy energy barrier is 0.8 eV. 
The major axis of the nanomagnet is $a$ = 300 nm
and the minor axis is $b$ = 50 nm to cause this shape anisotropy energy barrier.
The magnitude of the in-plane spin-polarized current generating spin-transfer-torque is 523 $\mu$A and the 
spin polarization of the current is 80\%. (a) Distribution of switching delay:  mean value = 2.1 ns, standard deviation = 0.48 ns, and (b) 
distribution of energy dissipation:  mean value = 4.1$\times 10^5$ kT [at room temperature], standard deviation = 2.6$\times 10^4$ kT [at room temperature].}
\end{figure}

Fig.~\ref{fig:thermal_distribution_delay_energy_cobalt} shows the 
room temperature (300 K) distributions of switching delay and energy dissipation, respectively, when both spin-transfer torque and thermal torque 
act on a nanomagnet with
in-plane shape anisotropy energy barrier of 0.8 eV.
The spin-transfer torque is due to an in-plane current of 523 $\mu$A with 80\% spin polarization.
The saturation magnetization $M_s=4.09\times10^5$ A/m. Here, 10,000 trajectories were simulated to compute these distributions.
We assume that when a spin-polarized current is applied to initiate switching, the magnetization vector
starts out from near the south pole ($\theta \simeq 180^\circ$) with a certain ($\theta_{init}$,$\phi_{init}$) picked from the initial angle distributions at 300 K. Switching is deemed to have completed when $\theta(t)$ becomes equal to or
less than 4.5$^{\circ}$. 
Note that both distributions are asymmetric about the mean. Regrettably, the long-delay tail extends much farther than the short-delay tail. The high-delay tail is associated 
with those switching trajectories that start very close to $\theta = 180^\circ$ which is a stagnation
point. In such trajectories, the starting torque is vanishingly small, which makes the switching 
sluggish at the beginning. During this time, switching also becomes susceptible to backtracking 
because of thermal fluctuations, which increases the delay further. The distribution of energy dissipation is complex, which shows multiple peaks in the distribution. The distribution is due to the variation in internal energy dissipation, $E_d$ (see Fig. 6 in the main paper).

\section{The benefits of switching with a larger switching current $I_s$}

\begin{figure}[htbp]
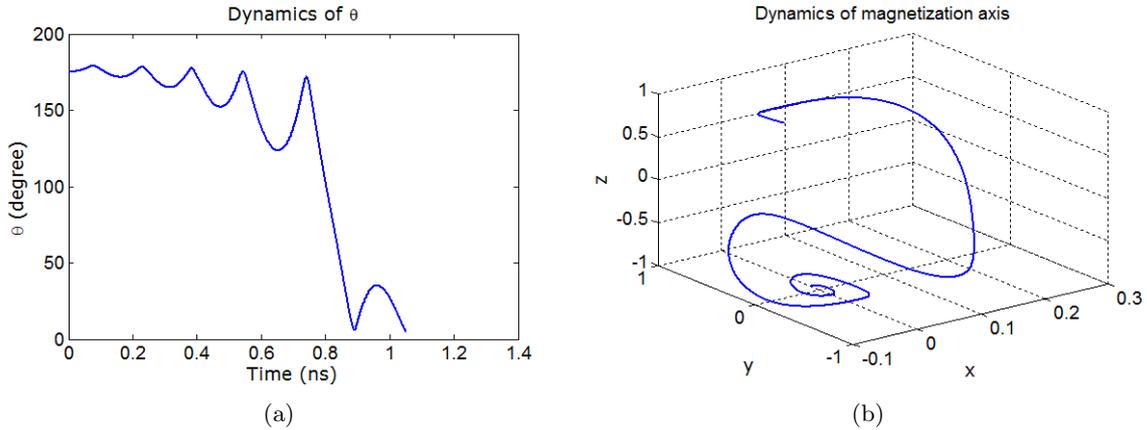

\centering
\subfigure[]{\label{fig:theta_dynamics_cobalt_Ms_8e5_Istt_2mA}\includegraphics[width=3in]
{theta_dynamics_cobalt_Ms_8e5_Istt_2mA}}
\subfigure[]{\label{fig:magnetization_dynamics_cobalt_Ms_8e5_Istt_2mA}\includegraphics[width=3in]
{magnetization_dynamics_cobalt_Ms_8e5_Istt_2mA}}
\caption{\label{fig:dynamics_cobalt_Ms_8e5_Istt_2mA} Switching dynamics of the magnetization vector in a nanomagnet of
major axis $a$ = 150 nm, minor axis $b$ = 100 nm, and $M_s = 8\times10^5$ A/m, and in-plane shape 
anisotropy energy barrier of 0.8 eV. This simulation does not consider any thermal fluctuations during the switching, however, the initial orientation of the magnetization is assumed to be $\theta_{init} = 175.5^{\circ}$ and $\phi_{init} = 90^{\circ}$ (thermally mean values for 300 K) to avoid the stagnation point exactly along the easy axis. Switching is caused by spin-transfer torque induced with
an in-plane current of 2 mA with 80\% spin polarization. (a) polar angle $\theta(t)$ versus time, 
and (b) the trajectory traced 
out by the tip of the magnetization vector during switching. Switching delay and energy dissipation are 
1.05 ns and $1.25\times10^7$ $kT$ [at room temperature], respectively.}
\end{figure}

\begin{figure}[htbp]
\centering
\subfigure[]{\label{fig:theta_dynamics_cobalt_Ms_8e5_Istt_4mA}\includegraphics[width=3in]
{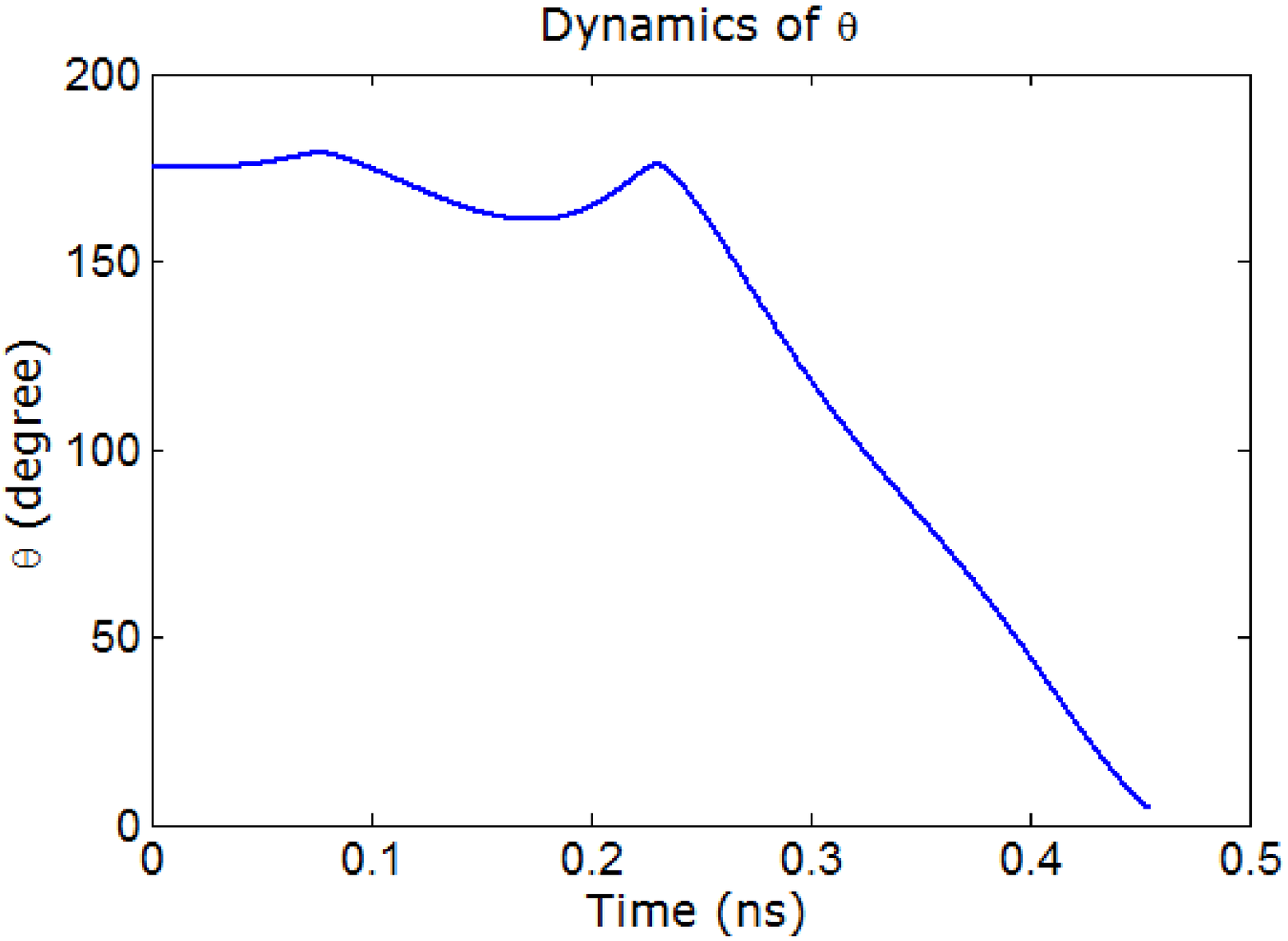}}
\subfigure[]{\label{fig:magnetization_dynamics_cobalt_Ms_8e5_Istt_4mA}\includegraphics[width=3in]
{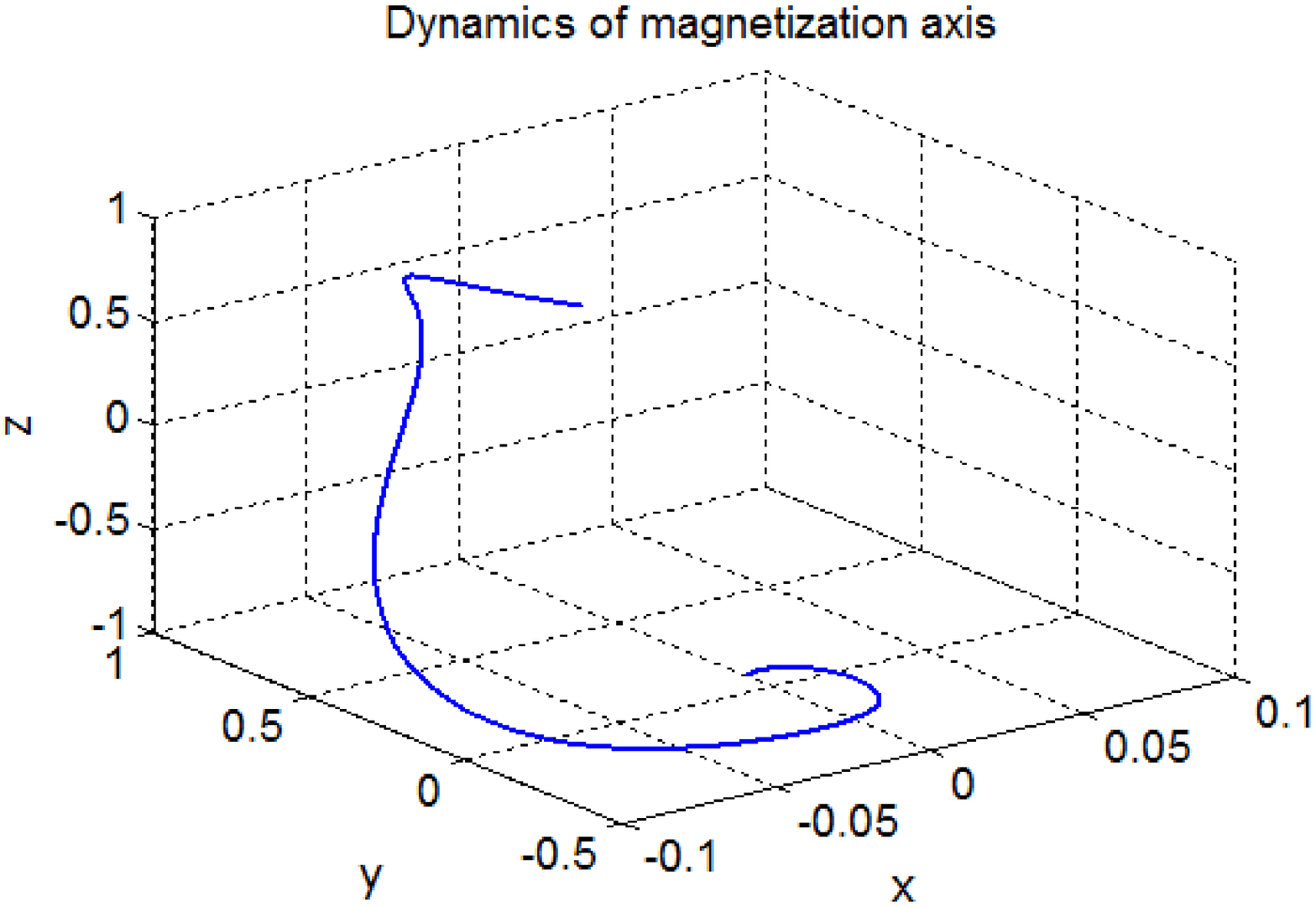}}
\caption{\label{fig:dynamics_cobalt_Ms_8e5_Istt_4mA} Switching dynamics of the magnetization vector in a nanomagnet of
major axis $a$ = 150 nm, minor axis $b$ = 100 nm, and  $M_s = 8\times10^5$ A/m. 
The resulting in-plane shape anisotropy energy barrier is 0.8 eV. 
This simulation does not consider any thermal fluctuations during the switching, however, the initial orientation of the magnetization is assumed to be $\theta_{init} = 175.5^{\circ}$ and $\phi_{init} = 90^{\circ}$ (thermally mean values for 300 K) to avoid the stagnation point exactly along the easy axis. Switching is caused by spin-transfer torque induced with
an in-plane current of 4 mA with 80\% spin polarization. (a) polar angle $\theta(t)$ versus time, and (b) the trajectory traced 
out by the tip of the magnetization vector during switching. Switching delay and energy dissipation are 455 ps and $2.2\times10^7$ $kT$ [at room temperature], respectively.}
\end{figure}

It is obvious that increasing switching current (and hence energy dissipation), while keeping saturation magnetization and shape anisotropy 
energy barrier constant, will result in faster switching. In order to illustrate this point, we do not consider any thermal fluctuations 
during the switching to avoid complications; however, the initial orientation of the magnetization is assumed to be $\theta_{init} = 175.5^{\circ}$ and $\phi_{init} = 90^{\circ}$ 
 to avoid the stagnation point exactly along the easy axis. Fig.~\ref{fig:dynamics_cobalt_Ms_8e5_Istt_2mA} shows the switching dynamics 
for this case when switching is caused by spin-transfer-torque induced with an in-plane switching current $I_s$ of 2 mA 
possessing 80\% spin polarization. 
The saturation magnetization $M_s=8\times10^5$ A/m. The switching occurs in $\sim$1 ns, while dissipating 1.25$\times$10$^7$ $kT$ [$T$ = 300 K] of
energy. If we increase $I_s$ to 4 mA (Fig.~\ref{fig:dynamics_cobalt_Ms_8e5_Istt_4mA}) while keeping everything else the same, 
the switching delay $\tau$ drops to 455 ps, 
but of course the energy dissipation 
(dominated by $I_s^2R \tau$) increases by 75\%. 
What Figs.~\ref{fig:dynamics_cobalt_Ms_8e5_Istt_2mA} and \ref{fig:dynamics_cobalt_Ms_8e5_Istt_4mA}  show is that the reason a higher switching current decreases switching delay is because it suppresses the ripples in the transient 
 dynamics of the magnetization vector. The ripples are caused by complicated precessional 
 motion of the magnetization vector shown in Fig.~S\ref{fig:magnetization_dynamics_cobalt_Ms_8e5_Istt_2mA}. A larger $I_s$ exerts a larger 
 spin-transfer torque on the magnetization vector that suppresses its errant precessional motion and brings about the switching faster.
 The important point is that since one can always decrease 
 switching current (and hence energy dissipation) by sacrificing switching speed, it is imperative to keep the thermal mean of 
the switching delay $\langle \tau \rangle$ constant 
 when studying how the switching current $I_s$ scales with saturation magnetization $M_s$. 

 
Increasing the switching current, while keeping everything else constant, also decreases the spread 
(variance) in switching delay caused by thermal fluctuations at non-zero temperatures. Monte Carlo simulations show that 
the standard deviation in switching delay 
decreases  from 0.23 ns to 0.14 ps at room-temperature when the switching current is increased from 2 mA to 4 mA. This 
happens because the larger switching current suppresses precessional motion of the magnetization vector 
that subjects the switching dynamics to greater variability 
in the presence of thermal fluctuations.
A smaller standard deviation in the switching delay helps increasing the operational clock-frequency and facilitates clock synchronization on a chip. Thus, expending more energy to switch (larger $I_s$) bears some distinct advantages in memory and logic applications, but of course at the cost of increased 
energy dissipation.

\begin{figure}
\centering
\includegraphics[width=3in]{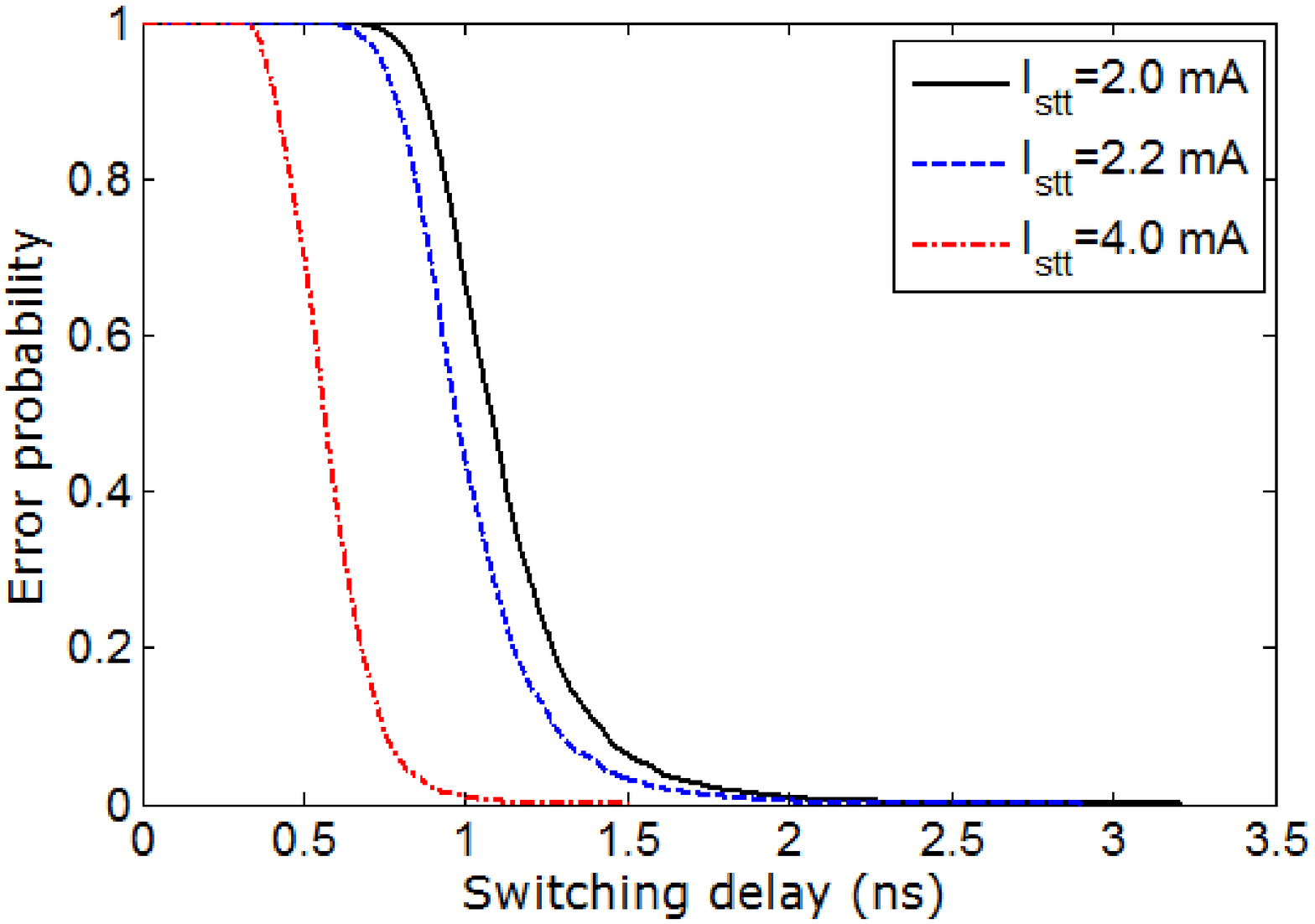}
\caption{\label{fig:error_probability_switching_delay} Dynamic switching error probability as a function of switching delay for different switching currents at 
$T$ = 300 K. The nanomagnet has dimensions of $a$ = 150 nm, $b$ = 100 nm, and the saturation magnetization is 
$M_s = 8\times10^5$ A/m. The resulting in-plane shape anisotropy energy barrier is once again 0.8 eV. 
Faster switching with a constant current increases the likelihood of error (failure to switch), but at a fixed speed of switching,
the error probability decreases with increasing switching current.}
\end{figure}

In Fig.~\ref{fig:error_probability_switching_delay}, we plot the dynamic switching error probability as a function
of switching delay $\tau$ for different switching currents $I_s$
in the presence of thermal fluctuations at room temperature. Switching {\it error} occurs when the 
magnetization vector sets out from its initial stable orientation along the easy axis towards the intended stable orientation
diametrically opposite, but fails to 
reach the latter within the stipulated delay $\tau$ because of thermal agitations.
For a fixed switching delay, a higher current ensures a higher probability 
of successful switching because it provides a stronger spin-transfer-torque
that overwhelms the thermal torque and 
suppresses the errant precessional motion of the magnetization vector that increases the chances of 
backtracking and the resulting switching failure. For a fixed switching current, the error probability increases with the speed 
of switching. This 
is somewhat obvious; if we allot less time for the switching to complete, it is less likely that switching 
will complete in that allotted time.

\section{Switching delay-energy as a function of $M_s$}

\begin{figure}[htbp]
\centering
\includegraphics[width=3in]{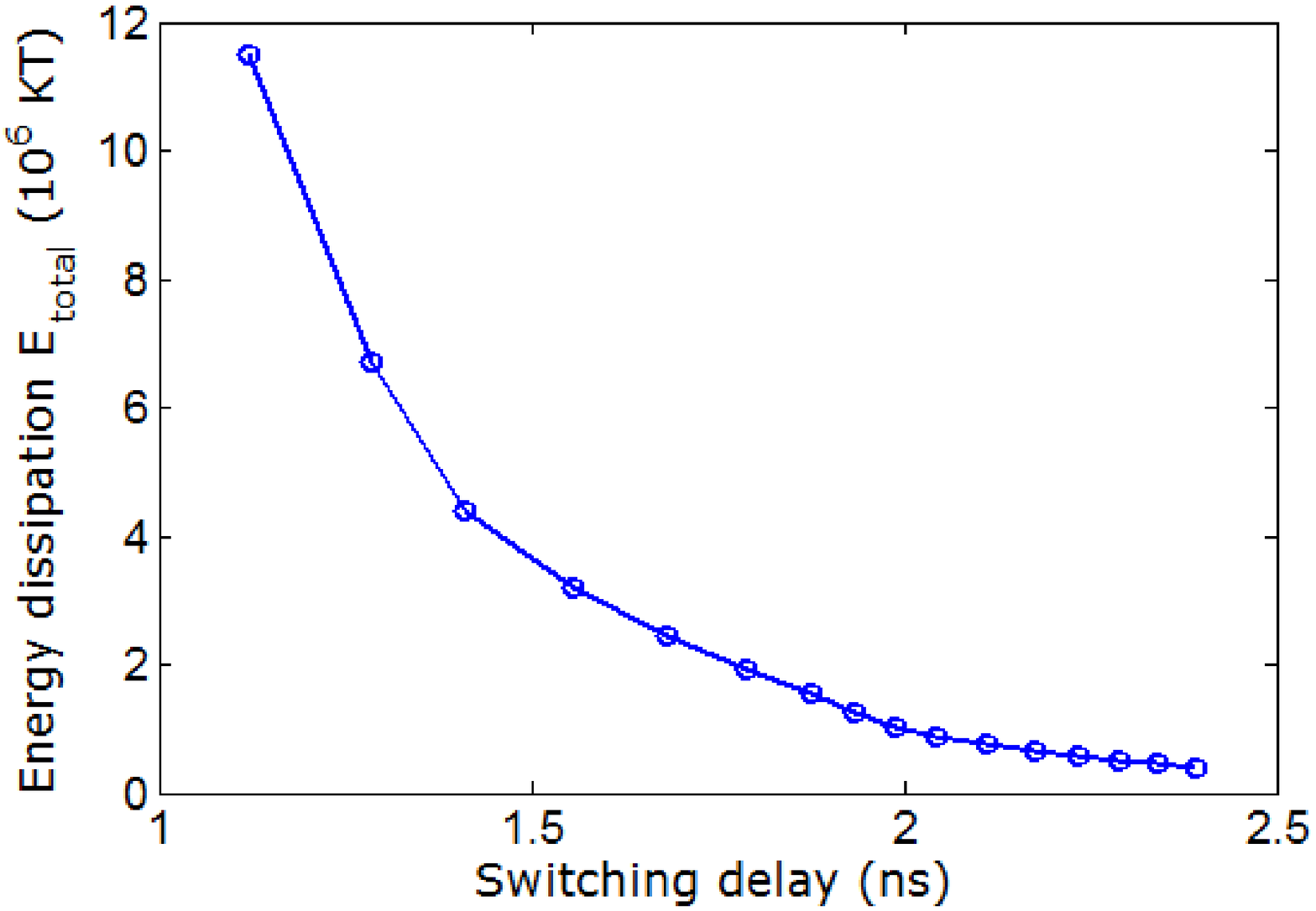}
\caption{\label{fig:delay_energy_thermal_mean} Thermal mean of the total energy dissipation versus thermal mean of the switching delay as a function of saturation magnetization $M_s$ ($4.09 \times 10^5 - 8 \times 10^5$ A/m). The square law scaling of switching current $I_s \propto M_s^2$ is used. A lower $M_s$ incurs less energy dissipation but switching delay is increased.}
\end{figure}

Fig.~\ref{fig:delay_energy_thermal_mean} shows the switching delay-energy trade-off as a function of saturation magnetization $M_s$. The plot shows a clear trade-off between the switching delay and energy dissipation as $M_s$ is varied. As $M_s$ is lowered, less energy is dissipated, however, switching delay is increased. The square law scaling of switching current $I_s \propto M_s^2$ is used. As discussed in the main paper, with lower values of $M_s$, a higher magnitude of current (than that of using the square law $I_s \propto M_s^2$) can reduce the switching delay to some smaller value corresponding to a higher value of $M_s$, while dissipating lesser energy than that of the high-valued $M_s$.

